\documentclass[12pt]{article}

\usepackage{multirow}
\usepackage{calc}
\usepackage{amsmath,amssymb,amsthm,amscd}
\numberwithin{equation}{section}
\usepackage[bf]{caption}
\usepackage{longtable}
\usepackage{array}
\usepackage{enumerate}
\usepackage{float}
\usepackage{subfig}
\usepackage{multicol}
\usepackage{multirow}
\usepackage{epsfig}
\usepackage{graphicx}
\usepackage{pict2e}
\usepackage{color}
\usepackage{authblk}
\usepackage{cite}
\usepackage[all]{xy}
\usepackage[width=1.09\textwidth]{caption}
\usepackage{bbm}

\usepackage{amsmath,amsfonts,amssymb}
\usepackage{bbm}

\usepackage[utf8]{inputenc}
\usepackage[T1]{fontenc}
\usepackage[english]{babel}
\usepackage{xspace}
\usepackage{pifont}

\usepackage[colorlinks=false, linkcolor=blue, citecolor=blue, urlcolor=blue]{hyperref}

\begin{document}


\title{Tracing symmetries and their breakdown through phases of heterotic (2,2) compactifications}
\author[a]{\bf Michael~Blaszczyk\thanks{blaszczyk@uni-mainz.de}}
\author[b]{\bf Paul-Konstantin~Oehlmann\thanks{oehlmann@th.physik.uni-bonn.de}}
\affil[a]{\it Johannes-Gutenberg-Universit\"at,Staudingerweg~7,~55099~Mainz,~Germany}
\affil[b]{\it Bethe~Center~for~Theoretical~Physics, Physikalisches~Institut~der~Universit\"at~Bonn,
Nussallee~12,~53115~Bonn,~Germany}
\maketitle
\abstract{\noindent We are considering the class of heterotic $\mathcal{N}=(2,2)$ Landau-Ginzburg orbifolds with 9 fields corresponding to $A_1^9$ Gepner models. We classify all of its Abelian discrete quotients and obtain 152 inequivalent models closed under mirror symmetry with $\mathcal{N}=1,2$ and $4$ supersymmetry in 4D. We compute the full massless matter spectrum at the Fermat locus and find a universal relation satisfied by all models. In addition we give prescriptions of how to compute all quantum numbers of the 4D states including their discrete R-symmetries. Using mirror symmetry of rigid geometries we describe orbifold and smooth Calabi-Yau phases as deformations away from the Landau-Ginzburg Fermat locus in two explicit examples. We match the non-Fermat deformations to the 4D Higgs mechanism and study the conservation of R-symmetries. The first example is a $\mathbb{Z}_3$ orbifold on an E$_6$ lattice where the R-symmetry is preserved. Due to a permutation symmetry of blow-up and torus K\"{a}hler parameters the R-symmetry stays conserved also smooth Calabi-Yau phase. In the second example the R-symmetry gets broken once we deform to the geometric $\mathbb{Z}_3 \times \mathbb{Z}_{3,\text{free}}$ orbifold regime.
}
\clearpage

\tableofcontents
\clearpage


\section{Introduction}
\label{sec:Introduction}
One of the greatest features of gauged linear sigma models (GLSMs) is the possibility to describe very different types of string compactifications
in a unified way. In the two dimensional theory the Fayet Illiopolus (FI) terms play the role of K\"{a}hler parameters of the target space geometry\cite{Witten:1993yc}. By tuning those to negative values the associated dual cycles shrink to zero size resulting in partially or fully singular phases such as orbifolds or Landau-Ginzburg models \cite{NibbelinkGroot:2010wm,Aspinwall:2012qca,Blaszczyk:2011hs}. Hence GLSMs provide a very nice tool for string construction and deep insights into their relations. However not every phase has a nice conformal field theory (CFT) descrpition which makes the computation of the spectrum in the 4D theory hard for a generic phase. In the orbifold regime the full CFT is known and the full matter spectrum and all symmetries can be obtained whereas this is not the case in the smooth Calabi-Yau regime.\\\\
With the virtue of having a UV complete theory at hand it is a great desire to classify all the possible compactification spaces and obtain their properties and possible connections. Hence a lot of progress has been made to classify those spaces within the context of Landau-Ginzburg models \cite{LUTKEN1988152,Font:1989rs,Fuchs:1989yv,Lynker:1990ez } orbifolds \cite{Fischer:2012qj} as well as in smooth Calabi-Yaus described as hypersurfaces \cite{Kreuzer:2000xy} complete intersections \cite{Altman:2014bfa} in projective varieties.\\\\
In accordance to the insights into the string geometry of the compactifications spaces it was always a main effort to connect string theory with familiar particle physics. Here basically all phases have been explored for model building, i.e. Gepner model constructions \cite{Blumenhagen:1995tt,GatoRivera:2009yt}, orbifolds \cite{Lebedev:2006kn,Blaszczyk:2009in,Pena:2012ki,Pena:2013lla,Nilles:2014owa} as well as smooth compactifications \cite{Braun:2005nv,Anderson:2012yf,Anderson:2013xka}. In particular Gepner and orbifold models seem to be a good starting point for model building as they have a lot of additional i.e.\ discrete symmetries that are useful for models of particle physics. Those symmetries have a very nice geometrical interpretation: Non-Abelian Flavor symmetries arise from permutation symmetries of orbifold singularities \cite{Kobayashi:2006wq,Nilles:2012cy} and discrete R-symmetries are remnants from the rotations of the 10D Lorentz symmetry preserved by the orbifold action. Hence orbifold spaces are the natural geometry where one should expect R-symmetries \cite{Bizet:2013gf,Bizet:2013wha,Nilles:2013lda}. On the other hand using the CFT techniques it is  hard to obtain those symmetries in the case of non-factorized torus lattices or in cases when freely acting involutions are modded out as well\cite{Bizet:2013wha}.\\
However starting from the orbifold gauge group and matter content is usually much larger then the ones of the MSSM and hence they have to be sufficiently be reduced. Those reductions often times correspond to resolutions of the orbifold singularities towards smooth Calabi-Yau phase. Hence it might also be natural to consider smooth geometries right from the start. These approaches have the benefit of having huge amounts of geometric data sets to be explored in addition to the flexibility in choosing a poly stable vector bundle over it. However in those cases possible helpful stringy effects might not be visible in the SUGRA approximation. Especially additional symmetries that can control proton decay inducing operators rely on continuous Abelian symmetries \cite{Buchbinder:2014qda} as the discrete ones are much less understood and there is no reason to expect a meaningful R-symmetry.\\
Here the orbifold picture can be helpful in uncovering such symmetries as discrete remnants of Higgsed symmetries that correspond to the blow-up modes. In some cases there might even be R-symmetries when the singularities are blown up in a symmetric way such that rotational symmetries might stay preserved \cite{Ludeling:2012cu,Nilles:2013lda}.\\\\
Landau-Ginzburg orbifold models were first examples where the phenomenon of mirror-symmetry was observed and physically explained by a simple sign change of the (left-moving) U(1) R-symmetry of the (2,2) CFT \cite{Greene:1990ud}. However in the target space geometry the effect is much more drastic and in particular in the smooth geometric regime described by non linear sigma models (NLSMs), the interpretation of complex structure and K\"{a}hler deformation gets interchanged. Mirror symmetry has lead to deep insights into the counting of rational curves in smooth Calabi-Yau \cite{Candelas199121} and the computation of physical Yukawa couplings in terms of the mirror dual Landau-Ginzburg description \cite{Chun:1989se}.\\
The benefit of the Landau-Ginzburg  description in particular for (2,2) models lies in  the strong world sheet symmetries that ensure a save running towards a CFT in the IR to a minimal model CFT \cite{Witten:1993jg}. Landau-Ginzburg models come with their own set of techniques that makes it possible to calculate the full massless spectrum for arbitrary superpotentials \cite{Kachru:1993pg} far beyond the chiral-chiral and chiral- anti-chiral ring.\\
Indeed by the GLSM intuition we have seen that non zero FI parameters corresponds to Higgsings of the 4D theory and hence one expects the Landau-Ginzburg phase in that sequence to have the highest amount of symmetry. Hence if we have control over the Landau-Ginzburg (LG) phase of an orbifold or smooth phase we can track the origin of possible (discrete-)remnant symmetries found in the orbifold phase. These questions might be of particular interest as they can uncover the point in the moduli space where the discrete symmetries becomes local as expected by general arguments of quantum gravity \cite{BANKS198990}.\\\\
Especially geometries that do not have complex structure deformation are particularly interesting as their whole moduli space is spanned by the K\"{a}hler moduli only. In such cases the LG model of the mirror can actually describe the full deformation space completely by their polynomial deformation while it stays fixed at the zero volume locus. In such a case we can describe the full spectrum and its changes throughout all phases of the mirror dual GLSM. Finally we note that remarkable steps have been made towards the computation of the full massive and massless spectrum of the GLSM in an arbitrary phase. This is made possible by powerful techniques of supersymmetric localization of the one-loop partition function \cite{Benini:2013nda,Benini:2013xpa,Nibbelink:2014ula}. However we concentrate on the direct derivation of the spectrum using the methods of \cite{Kachru:1993pg} and the concrete computation of charges in this work.
\\\\
This paper is structured as follows: 
First we review the methods of how to calculate the full massless spectrum of Landau-Ginzburg models in Section \ref{sec:LGOs}. Having clarified those methods we 
introduce a way of how to compute the charges of {\it all} symmetries of these models, that include gauge charges and the ones of discrete R-and non-R symmetries. These techniques underlie the classification of all $A_1^9$ Landau-Ginzburg orbifold (LGO) models we present in Section~\ref{sec:Sec3} including the computation of the full spectrum at the Fermat point. We discuss the results of the scan and general features of the models.
In Section~\ref{sec:TargetSpace} we discuss two explicit examples in detail. We explicitly compute the charges under all symmetries and track their conservation/breakdown through various phases matching the LGO deformation perfectly with 4D effective theories. 
In Section \ref{sec:summary} we discuss and summarize our results. In Appendix \ref{app:LGOList} we list the defining properties of our classification, up to mirror symmtry and the full spectrum at the Fermat point.
\section{Landau-Ginzburg Orbifolds and Their Symmetries}
\label{sec:LGOs}
In this section we review the tools necessary to analyze the massless spectrum of Landau-Ginzburg orbifold models. Although we stick to a very specify class of models this review is 
general and applicable to other cases as well. At first we review the methods that are needed to calculate the massless spectrum. In the second part we give the methods needed to calculate all charges of the spectrum under all discrete and continuous symmetries.
\subsection{Landau-Ginzburg models and their spectrum}
In the following we review the methods to calculate the full massless spectrum of Landau Ginzburg orbifolds using techniques developed in \cite{Kachru:1993pg}. For another review also see \cite{Aspinwall:2010ve}.\\
We are considering $\mathcal{N}=(2,2)$ supersymmetric two dimensional field theories with chiral superfields $\Phi^i$ charged under left-and right-moving R-symmetries $(q_- , q_+)$
such that the Landau-Ginzburg superpotential $\mathcal{W}$ is a homogeneous function of degree one. Left-and right-moving R-charges of the superfields $\Phi^i$ are the same and given by $q_-^i = q_+^i = \alpha^i$. In the infra red the Landau-Ginzburg model flows to a minimal model CFT \cite{Witten:1993jg} with central charge 
\begin{align}
\label{eq:centralc}
c = 3 \sum_i (1 - 2\alpha^i) \, .
\end{align}
We want to describe $c=(9,9)$ compactifications and focus on the subclass with nine chiral fields with R-charges $\alpha^i = 1/3$ in the rest of this work.
To complete this theory to a critical heterotic string theory we have to add the following field content:
\begin{itemize}
\item four light-cone gauged left moving bosons and four right-moving $\mathcal{N}=(1,0)$ multiplets to provide non-compact 4D Minkowski space.
\item ten left moving Mayorana Weyl fermions $\lambda^I$ contributing an SO(10) symmetry gauge symmetry.
\item eight left chiral Bosons compactified on an even and self-dual torus providing an $\rm{E}_8$ gauge symmetry.
\end{itemize}
Space-time supersymmetry is guaranteed by GSO projection onto states with integral left-and right moving R-charge $q_-$ and $q_+$ \cite{Gepner:1987qi}. In addition to space-time supersymmetry the GSO projection ensures the lift of the left moving current $J_-$ and SO(10) to $E_6$. The GSO projection leaves only states with
\begin{align}
\label{eq:GSO}
\hat{g} :=  e^{- \pi i \left( J_- + F\right)}  = 1\,,
\end{align}
with $F$ being the left-and right-fermion number of the oscillator operators. Since $\alpha^i$ is an R-charge the bosonic and fermionic fields in an $(2,2)$ world sheet superfield have different charges summarized in Table \ref{tab:LgorCharges}.
\begin{table}[h!]
\begin{center}
\begin{tabular}{|c|c|c|c|c|}\hline
charge & $\phi^i$ & $\psi^i$ & $\overline{\phi^i}$ & $\overline{\psi^i}$ \vphantom{\Big|}\\ \hline
$q_{-}$ & $\alpha^{i}$ & $\alpha^{i} - 1$ & $-\alpha^{i}$ & $1-\alpha^{i}$ \\ \hline
$q_{+}$ & $\alpha^{i}$ & $\alpha^{i}$ & $-\alpha^{i}$ & $-\alpha^{i}$ \\ \hline
\end{tabular}
\caption{\label{tab:LgorCharges}Left- and rightmoving R-charges of bosonic and fermionic components of superfields.}
\end{center}
\end{table}
Since we have rational R-symmetry charges $\alpha^i$ the exponent of the GSO operator will  give non-trivial projection constraints on the
fields but there is a power $N$ such that
\begin{align}
\hat{g}^{N} = {\rm id} \, .
\end{align}
By consistency we have to supplement the orbifold action with the addition of $N-1$ twisted sectors
where the world sheet fields close upon the twisted boundary conditions of $\hat{g}$. Especially for our choice of $\alpha_i=\frac13$ we always have to
consider at least six twisted sectors.\footnote{Note that $g^{\frac{1}{\alpha}}$ results in a $-1$ phase which results in the NS boundary conditions on fermions. Hence we have effectively mixed these sectors into the orbifold identification.
}\\\\ In addition to the R-symmetry we can impose additional discrete non-R symmetries 
to the Landau-Ginzburg theory that restrict the superpotential $\mathcal{W}$ and leads to further linear independent twistings. From the perspective of the two dimensional field theory those discrete symmetries can often be obtained as discrete remnants of gauged U(1) symmetries that originate from a GLSM description. Indeed many of our models can be described as a Landau-Ginzburg orbifold phase of such models.\\
In general the addition of $N$ discrete symmetries $\mathbb{Z}_{n_j}^{(j)}$, $j = 1 , \ldots,N$ leads to additional  projections and additional twisted sectors.
Hence each additional $\mathbb{Z}^{(j)}_{n_j}$ factors represents another orbifold taken from the original theory. 
In total there can be $\prod^N_{j=0} (n_j)-1$ additional twisted sectors which can easily exceed $\mathcal{O}(100)$.
In the following we denote a twisted sector by the column vector $(k_0; k_1,\ldots,k_N)$ where we highlight the R-symmetry twist by the first entry $k_0$. But note that the additional quotients we take are Abelian non-R symmetries and hence all fields within one chiral multiplet $\Phi^i$ have the same charge under them. Collecting the contributions of all twsitings in a $(k_0; k_1,\ldots,k_N)$ twisted sector we obtain for bosonic (fermionic) coordinates $\phi^i$ ($\psi^i$) the total oscillator shift of $\nu^i$ ($\widetilde{\nu}^i$) given as
\begin{align}
\nu^i & = \frac{k_0 \alpha^{i}}{2} + \sum^N_{j=1} k_j Q^i_{j}   &\text{ mod } 1  \quad \text{ with } & 0 \leq \nu^i \leq 1 \\
  \widetilde{\nu}^i & = \frac{k_0 (\alpha^{i}-1)}{2} + \sum^N_{j=1} k_j Q^i_{j}& \text{ mod } 1  \quad \text{ with } &-1 \leq \widetilde{\nu}^i \leq 0
\end{align}
where the superfield $\Phi^i$ has a $\mathbb{Z}^{(j)}_{n_j}$ charge $Q^i_j$. Due to the twistings the vacuum acquires a non trivial energy contribution
\begin{align}
E_{\text{vac}} = \left\{      \begin{array}{ll} - \frac58 + \frac12 \sum_i \left(\nu_i (1-\nu_i)+ \widetilde{\nu_i}(1+\widetilde{\nu_i})\right) &  \text{for $k_0$ odd} \\ 0 & \text{for $k_0$ even}  \end{array} \right\}   \, .
\end{align}
and charges
\begin{align}
\label{eq:vaccharges}
q_{-,\text{vac}} =& \sum_i \left(  (\alpha_i-1)(\widetilde{\nu_i}-1)-\alpha_i (\nu_i - \frac12)                \right) \, ,\\
q_{+,\text{vac}} =& \sum_i \left(  \alpha_i(\widetilde{\nu_i}+\frac12)+ (\alpha_i-1) (-\nu_i + \frac12) \right)  \, , \\
Q_{j,\text{vac}} =& \sum_i  Q_j^i \left( \widetilde{\nu_i} - \nu_i - 1 \right) \,.
\end{align}
We construct a state by acting with oscillators on the vacuum to obtain $E=0$ states and impose the GSO and $\mathbb{Z}_{n_j}^{(j)}$ projections. Here we are sticking to the convention that negative oscillator frequencies have positive energy contributions, as in \cite{Kachru:1993pg}.
The $q_-$ and $q_+$ quantum numbers of a 4D state are given by the sum of the vacuum in eq. \eqref{eq:vaccharges} and oszillator contributions given in Table~\eqref{tab:LgorCharges}.\\
We iterate this procedure for each twisted sector and collect all states with same R-charge quantum numbers into vector spaces distinguished by left- and right-moving charge: $V_{(q_-,q_+)}$. \\
When we construct the states above we are particularly interested in states massless under the left- and right-moving Hamiltonian of the 2D theory
\begin{align}
2L_{\pm,0}= \{Q_\pm , \overline{Q}_\pm \} = 0 \, .
\end{align}
The supercharges are nil potent and hence we are looking for states in the cohomology of the $\overline{Q}_\pm$ operators. 
As the right moving part gives rise to space-time supersymmetry this operator is of particular importance for us.
The $Q_+$ operator commutes with $Q_-$ and hence does not change $q_-$ of a state by its action but raises $q_+$ by one unit. Hence $\overline{Q}_+$ acts as a map between the vector spaces that gives rise to the following complex
\begin{align}
...\stackrel{\overline{Q_{+}}}{\longrightarrow} V_{(q_- , q_+)} \stackrel{\overline{Q_{+}}}{\longrightarrow} V_{(q_- , q_+ +1)}  \stackrel{\overline{Q_+}}{\longrightarrow}.... \, .
\end{align}
where massless 4D states correspond to states in the $\overline{Q}_+$ cohomology
\begin{align}
H = \frac{\text{ker}(\overline{Q}_{+})}{\text{im}(\overline{Q_{+}})} \, ,
\end{align}
in the respective segment of the complex.
In principle one should also obtain the $\overline{Q}_-$ cohomology but this is already achieved by the GSO projection.
We construct the operator $\overline{Q}_+$ explicitly in terms of components of the $\Phi^i$ fields by integrating over its Noether current and obtain 
\begin{align}
\overline{Q}_+ = \int d \sigma \overline{\psi}^i_+ \partial_+ \phi^i + i \psi^i_- \partial_{\phi^i} \mathcal{W}^\prime \, ,
\end{align}
where the first part comes from K\"{a}hler potential and the second from the super potential. 
As explained in \cite{Kachru:1993pg} we can compute the cohomology of $\overline{Q}_+$ by looking at the K\"{a}hler potential and superpotential contribution independently. 
 The parts coming from the K\"{a}hler potential restricts us to consider states without $\psi_+$ modes and those that only depend 
holomorphically on the bosons $\phi$ as these have a non-vanishing (anti-)commutator and thus cannot be in the kernel of $\overline{Q}_+$.  
Hence with this restriction in mind we can reduce our considerations to the superpotential terms 
\begin{align}
\overline{Q}^\prime_+ =  \psi^i_- \partial_{\phi^i} \mathcal{W}^\prime\, .
\end{align}
The above computation can be extremely time consuming in particular when there are $\mathcal{O}(100)$ twisted sectors.
\\
Having computed left- and right-moving R-charges of every state that are in the $\overline{Q}_+$ cohomology we can identify their space-time properties.
As noted in the beginning only the SO(10) gauge symmetry is explicit as the rotational symmetry of the Mayorana-Weil fermions. The left-moving world sheet current however becomes a U(1) current in the four dimensional space that enhances the SO(10) to E$_6$.
 Hence an $\rm{E}_6$ representation can be identified by a collection of states and their $q_-$ charges according to their group
theoretical decomposition
\begin{align*}
\mathbf{78}& \rightarrow \mathbf{45}_0 \oplus \mathbf{16}_{-3/2} \oplus \overline{\mathbf{16}}_{3/2} \oplus \mathbf{1}_0 \, , \\
\mathbf{27}& \rightarrow \mathbf{16}_{-1/2} \oplus \mathbf{10}_1 \oplus \mathbf{1}_2 \, , \\
\mathbf{1} & \rightarrow \mathbf{1}_0 \, .
\end{align*}  
In a similar fashion the right-moving charge $q_+$ identifies the supersymmetric representation. 
This can in general be done by constructing the vertex operators corresponding to the space-time super fields and the SUSY generators \cite{Kachru:1993pg}. We do not review this construction here but state the result that a state with $q_+ = - \frac12$ is a left-chiral fermion and states with $q_+ =  -\frac32$ are gauginos in a vector multiplet. As the bosonic content of the theory is fixed by space-time supersymmetry it is sufficient to calculate the fermionic spectrum of the theory.\\\\
These are all the necessary steps that we need to consider in order to obtain the spectrum of a given Landau-Ginzburg orbifold model. Once again we clarify the input data necessary to fix a Landau-Ginzburg model completely:
\begin{enumerate}
\item {\bf Master Model:} Fix the superfield content and their R-charges constrained by \eqref{eq:centralc} to give a central charge $c=(9,9)$ in the IR.
\item {\bf Discrete Quotients:} Choose possible additional discrete quotients of the master model by an anomaly free charge distribution of the chiral
superfields.
\item {\bf Superpotential $\mathcal{W}$:} Choose a superpotential $\mathcal{W}$ as homogenous function in the fields consistent with all symmetries.
\end{enumerate}
In this work we have fixed the master model and have classified all discrete quotients and computed the whole spectrum for Fermat superpotentials that we give in Section~\ref{sec:Sec3}. In Section~\ref{sec:TargetSpace} we consider models where we deform away from the Fermat locus.
\subsection{Construction of target space symmetries}
\label{subsec:TargetSymmetries}
In the previous section we have shown how we can deduce the $E_6$ representation of the states from the charges of the left-moving U(1) symmetries.  
In this section we focus on the various other symmetries outside of $E_6 \times E_8$.\\\\
We start by constructing the space-time R-symmetry generator, generalizing methods of  \cite{Distler:1994hs}. The key ingredient to construct these generators is the observation that neither space-time supersymmetry nor E$_6$ symmetry is explicit. I.e. the $E_6$ gauginos are generically distributed among the first $k_0$ twisted sectors as:
\begin{center}
\begin{tabular}{|c|c|c|c|c|} \hline
State & $\mathbf{1}_0$ & $\mathbf{45}_0$ & $\overline{\mathbf{16}}_{-\frac32}$ & $\mathbf{16}_{\frac32} $\\ \hline
Sector& $(1;0,\ldots, 0)$ &$(1;0,\ldots, 0)$ & $(0;0,\ldots, 0)$ & $(2;0,\ldots, 0)$ \\ \hline
\end{tabular}
\end{center}
As all states belong to a vector multiplet we have to find a charge operator that gives a universal charge for all sub representations. Clearly this operator must be sensitive to the $k_0$ twist but also needs to be orthogonalized with respect to additional gauge symmetry generators. In the following we propose the following charge generator
\begin{align}
\label{eq:z3rsymm}
Q^{(M)}_R = 3k_0 - 2 q_-  + \sum_i n_i q_{i} \text{ mod } M  \,  \text{ with } n_i \in \mathbb{Z} \, ,
\end{align}
with $M$ being $ \frac{6}{\alpha}$. Our proposition for this generator differs by the one given in \cite{Distler:1994hs} by the correction terms in possible non-Abelian Cartan elements $q_i$. As gauginos of the Cartan U(1)'s are always in the $(1;0,\ldots, 0)$ sectors they all have R-charge $q_R = 3$. However in some cases additional gauginos in other twisted sectors may appear as the roots of an additional non-Abelian enhanced gauge symmetry and must also have the same R-charge as the Cartan generators. Hence the coefficients $n_i$ in \eqref{eq:z3rsymm} must be chosen to give a universal charge also to those states.
As the R-symmetry of the gauginos is fixed to be $3$ the 4D superpotential has R-charge $Q^{(M)}_R\left(\mathcal{W}_{4D}\right) = -6$ mod $M$.\\\\
Next we want to propose a method to calculate the the charges under the additional U(1) symmetries directly in terms of the discrete charge of the Landau-Ginzburg fields with Fermat superpotential. First it is readily checked that each world sheet chiral multiplet $\Phi^i$ automatically leads to a 4D gaugino state of the form 
\begin{align}
\left( \phi^{i}_{-\frac16} \overline\phi^{i}_{-\frac56} - 2 \psi^{i}_{-\frac13} \overline\psi^{i}_{-\frac23} \right)|1;0, \ldots, 0\rangle \,, i = 1 \ldots 9 \, ,
\end{align}
with $\phi_r^i, \psi_r^i$ being modes of the WS component fields, that generate nine U(1) currents. The trace of them is just the left-moving U(1) inside of $E_6$ which explains the first correction term in the R-symmetry generator eq. \eqref{eq:z3rsymm}. 
In analogy to \ref{eq:vaccharges} we define the charges of the vacuum world sheet bosons, fermions and their conjugates as
\begin{align}
\begin{tabular}{|c|c|c|c|c|c|}\hline
Operator& $|\text{vac} \rangle$ & $\phi^j$   & $\overline{\phi}^j$ & $\psi^j$   & $\overline{\psi}^j$ \\ \hline
Charge $q^i$ & $(\alpha^j -1)\cdot(\hat{\nu}^i + \frac12) - \alpha^i \cdot (\nu^i - \frac12)$ & $\alpha^i \delta_{i}^j$ & $-\alpha^i\delta_{i}^j$ & $(\alpha^i-1)\delta_{i}^j$ & $(1-\alpha^i)\delta_{i}^j$ \\ \hline
\end{tabular} \, .
\end{align}
In the cases when additional gauginos appear in other than the first twisted sector they give rise to massless $W$ bosons of the enhanced gauge symmetry. In this case we can calculate the corresponding gauge enhancement by finding the Cartan charge that are given by an appropriate linear combination of the above Cartan charge operators.
Explicit examples of SU(3) enhancements can be found in Figure \ref{fig:vaccharges}.
Finally there are additional Abelian discrete $\mathbb{Z}^{(j)}_{n_j}$ symmetries provided by the additional twisted sector quantum numbers. Note that also
that the discrete charges can be corrected by U(1) generators as well if non-Abelian enhancement occurs
\begin{align}
Q^{n_j}_j = k_j +  \sum_i n_i q_{i} \text{ mod } n_j \, .
\end{align}
By the same argument by which we have an R-symmetry generator it is clear that the above generator cannot be an R-symmetry because Cartan gauginos are always 
in the $(1;0,\ldots, 0)$ sector and hence must have trivial charge under the same and conversely also the 4D superpotential must be trivially charged under those.
\section{Classification of $A^9_1$ models}
\label{sec:Sec3}
There is a vast literature on classifications and constructions of $\mathcal{N}=(2,2)$ Landau Ginzburg orbifold models \cite{GEPNER1988757,LUTKEN1988152,Lynker:1990ez,Fuchs:1989yv,Font198979,Klemm:1992bx,Schellekens:1989wx} as well as for (0,2) models \cite{Font:1989rs,Blumenhagen:1996gz}  including computation of the chiral rings. We concentrate on the classification of (2,2) models with nine
chiral fields of R-charge $\alpha=\frac13$ and all of their inequivalent Abelian discrete $\mathbb{Z}_3^N$ quotients. Using the methods described in the previous section we complete the existing literature by the missing models and the full amount of uncharged vector and singlet states with respect to the $E_6$ gauge factors at the Fermat point.
Having found the enhanced symmetry sector, we can compute the full 4D symmetry group
and their charges that we focus on in Section \ref{sec:TargetSpace}. Moreover we find that the full set of 152 inequivalent models is closed under mirror symmetry. Finally we find that at the Fermat point all of the models obey the relation 
\begin{align}
N_{\text{Add-S}}-3N_{\text{Add-V}}=(4-\mathcal{N})\cdot 76 \, ,
\end{align}
where $N_{\text{Add-S}}$ and $N_{\text{Add-V}}$ count the amount of $\mathcal{N}=1$ chiral superfields and vectorfields uncharged under $E_6 \times E_8$
as well as the total amount of supersymmetries $\mathcal{N}$ of the model. We comment on this relation at the end of this section. 
\subsection{Fermat Classification and Mirror Symmetry}
First we note that the amount of discrete quotient factors is bounded by the constraint of inequivalent charges assignments to the nine superfields that become redundant as soon as there are more than nine $\mathbb{Z}_3$ discrete quotient factors. Consistency of the symmetries requires the discrete charge assignments to be anomaly free and hence they have to sum up to zero (mod 3) for every discrete factor which fixes w.l.o.g. the charge of one superfield uniquely. Moreover we can always rotate one linear combination into the R-symmetry to give one field trivial discrete charges. \\
 Thus the upper bound is fixed by at most seven $\mathbb{Z}_3$ quotients that we can take. By the above arguments the model with seven quotient has, up to redefinitions, a unique charge assignment given by
\begin{align}
\label{eq:Z37model}
Q_j (\Phi^i) &= \delta_{ij} \, \quad \text{ with }\quad j = 1,\ldots, 7 \, , \\
Q_j(\Phi^8) &=-1 \, \quad \text{ and }\quad Q_j(\Phi^9) =0\, . 
\end{align}
Summarizing we fix a model by a set of nine dimensional charge vectors $Q_j$ with $j=1\ldots N$. 
We then classify all models by systematically writing down all inequivalent charge assignment for a fixed amount of quotients $\mathbb{Z}_3^N$ with $N < 7$.
 Here two assignments are equivalent if they are related by charge redefinitions or permutations of the nine worldsheets fields $\Phi^i$.\\\\
{\bf  Summary of the Classification}\\\\
In total we find 152 inequivalent models. Depending on the amount of discrete quotient factors we summarize the number of inequivalent models in Table \ref{tab:Modelnumber}.
\begin{table}[t!]
\begin{center}
\begin{tabular}{|c|cccccccc|}\hline
Discrete Quotients $\mathbb{Z}^N_3$ N: & 0 & 1 & 2 & 3 & 4 & 5 & 6 & 7 \\ \hline
Inequivalent Models & 1 & 5 & 21 & 49 & 49 & 21 & 5 & 1 \\ \hline
\end{tabular}
\caption{\label{tab:Modelnumber}The summary of inequivalent models per number of $\mathbb{Z}_3$ quotients.}
\end{center}
\end{table}
The reflexion symmetry observed in Table \ref{tab:Modelnumber} is a consequence of Green-Plesser mirror symmetry \cite{Greene:1990ud} of the different 
Landau-Ginzburg orbifold models. 
The Green-Plesser mirror map can be summarized by the dualization of the set of charge vectors in $H$. The dual set of charge vectors in $H^\circ$ is obtained by
\begin{align}
\label{eq:mirrormap}
H^\circ := \left\{Q \in \mathbb{Z}_3^7 \, | \,\langle Q , Q'\rangle = 0 \text{ mod } 3 \quad \forall \,  Q' \in H          \right\}\, .
\end{align}
This indeed specifies for each collection of charge vectors $H \subset \mathbb{Z}_3^N$ a dual model with charge vectors in the complement vector space $H^\circ \subset \mathbb{Z}_3^{7-N}$. This is indeed a duality due to $H^{\circ\circ} = H$. One example is the mirror of the 
$\mathbb{Z}^7_3$ model. This model has already the maximal quotient group and hence the charge vector that are orthogonal to those must be trivial and thus its mirror is the master model.\\
Indeed it is known that modding out the complement symmetry of a model results in a sign change of the left-moving R-symmetry on the CFT level \cite{Greene:1990ud}.
In the target space geometry of non linear sigma models (NLSM) mirror symmetry acts as a change of complex structure and K\"{a}hler deformations that is a change of $\mathbf{27}$ and $\overline{\mathbf{27}}$ E$_6$ charged states in the $\mathcal{N}=(2,2)$ compactification. The change of those representations is the same but in other phases there might be not necessarily a geometric interpretation associated to them.
 The results of our classification in terms of charge vectors and the spectrum at the Fermat point is summarized in Table \ref{fig:LGOmodels} in Appendix \ref{app:LGOList} up to mirror symmetry.\\\\
{\bf The Landau-Ginzburg Superpotential}\\\\
The generic superpotential for our models are cubic monomials and are given as a sum of Fermat type superpotential $\mathcal{W}_{\text{Fermat}}$ and a
deformation part $\mathcal{W}_{\text{Deformation}}$:
\begin{align}
\mathcal{W} =& \mathcal{W}_{\text{Fermat}} + \mathcal{W}_{\text{Deformation}} \, , \\
\mathcal{W}_{\text{Fermat}} =& \sum_{i=1}^9 \Phi^3_i \, \qquad \mathcal{W}_{\text{Deformation}}=\sum_{1\le i<j<k\le9} a_{ijk} \Phi_i \Phi_j \Phi_k
\end{align}
The superpotential of Fermat type is a cubic coupling involving the same superfields only and all nine monomials have to be present in order to create compact directions for all coordinates. Those terms are allowed and necessary for arbitrary additional quotients.\\
 Depending on the charge assignment of the fields under additional quotient factors also other cubic monomials can be allowed and give rise to deformation terms. These polynomial deformations can generically be understood as complex structure deformations and decrease the full residual symmetry of the Landau-Ginzburg superpotential which is maximal at the Fermat point $a_{ijk}\equiv0$. Clearly there are charge assignments that are rigid and do not allow for those terms
at all. We consider such geometries in Section~\ref{sec:TargetSpace}.
Generically the $E_6$ charged matter is independent of such deformations, however the amount of additional $E_6 \times E_8$ uncharged vectors and gauge singlets depends on them.\\
However in our classification we focused on the Fermat type superpotentials and computed the full spectrum using the methods of \cite{Kachru:1993pg} described in Section~\ref{sec:LGOs} where the symmetries are maximal.
\subsection{Features of the classification}
\begin{figure}[h!]
\centering
\begin{picture}(250,160)
  \put(0,0){\includegraphics[scale=1]{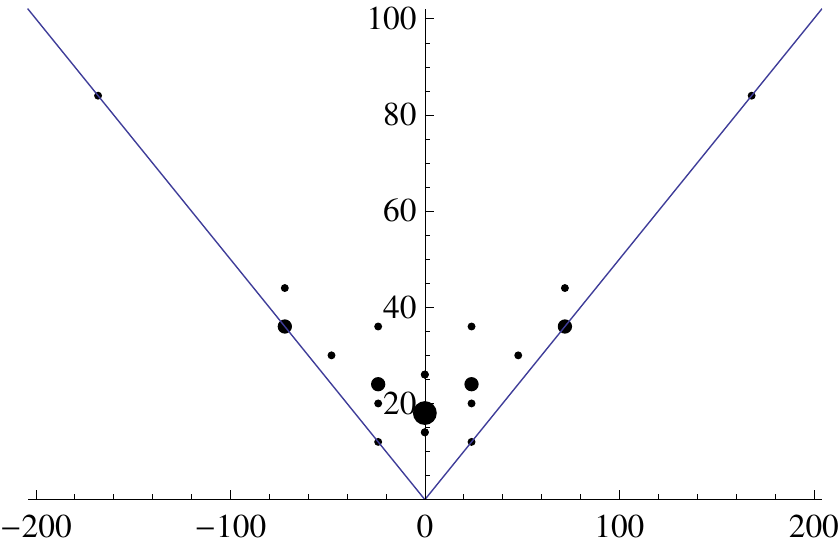}}
    \put(250,20){$\chi$}
    \put(140,150){$h^{(1,1)}+h^{(2,1)}$}
\end{picture}
\caption{\label{fig:ScanPlot}Summary of $\mathbf{27}$ and $\overline{\mathbf{27}}$ representations of $E_6$ of the $A_1^9$ LGO classification. In geometric analogy we plot the number of representations as hodge numbers. We plot their sum again the Euler number $\chi$.}
\end{figure}

In the following we have constructed the whole spectrum for all 152 inequivalent models at the Fermat point including gauginos and gravitinos as well as $E_6$ charged multiplets. We have summarized the spectrum of the $E_6$ charged matter in Figure \ref{fig:ScanPlot}. Note that we can identify states by the $U(1)_L$ charge as described in Section \ref{sec:LGOs}. In analogy to smooth heterotic compactifications using the standard embedding we identify the number of $\mathbf{27}$-plets as $h^{2,1}$ and $\overline{\mathbf{27}}$ as $h^{1,1}$ Hodge numbers, respectively. 
As anticipated the graph is fully mirror-symmetric realized by the reflection symmetry along the vertical-axis. Many models have the same Hodge numbers which is why only few spots in the plot are populated. Note that many of those models have an interpretation as the Landau-Ginzburg phase of an orbifold or smooth compactification. We comment on that relation in the following by subdividing the models
into three different classes:
\begin{enumerate}
\item {\bf $\boldsymbol{\chi=0}$ models:} Models along the y-axis have vanishing Euler numbers. Having vanishing Euler number is a necessary condition for higher supersymmetry as those theories are non-chiral in four dimensions. Indeed we find models of $\mathcal{N}=2$ and $\mathcal{N}=4$ SUSY besides $\mathcal{N}=1$ models. Throughout this work we present all representations in $\mathcal{N}=1$ language and hence higher SUSY multiplets have to be constructed from them accordingly. Higher SUSY models also have additional charginos as expected from their representation theoretic decompositions.\\
Among the $\mathcal{N}=2$ models those with $h^{(2,1)}=h^{(1,1)}=21$ are the most prominent example as they are the Landau-Ginzburg loci of $K3\times T^2$ compactification as their Hodge numbers can be computed as
\begin{align}
h^{2,1}(K3\times T^2) &= h^{1,1}(K3)\cdot h^{1,0}(T^2) + h^{2,0}(K3)\cdot h^{0,1}(T^2) = 21 \, , \\
h^{1,1}(K3\times T^2) &= h^{0,0}(K3)\cdot h^{1,1}(T^2) + h^{1,1}(K3)\cdot h^{0,0}(T^2) = 21 \, .
\end{align}
 All other $\mathcal{N}=2$ compactification have a $K3 \times T^2$ origin but with additional $\mathbb{Z}_3$ involutions modded out.
Similarly we find two different $\mathcal{N}=4$ models originating from $T^6$ compactifications at the Landau-Ginzburg locus, one with additional $32$ vectors and one with $86$. Indeed in both cases 8 of those additional Vector multiplets form the adjoint representation of SU(3) that enhances (together with the three $(\mathbf{27},\mathbf{3})$ reps) $E_6$ to the full $E_8$ as expected from a trivial torus compactification. The $T^6$ torus structure becomes even more visible by inspecting the residual vector states.\\ In the first case the 24 vector multiplets originate from the winding modes of the torus that become massless at the Fermat Landau Ginzburg locus. They correspond to the $SU(3)^3$ adjoints of the same Lie lattice that underlies the $T^6$ which becomes fully gauged at this point in the K\"{a}hler moduli space.\\ The situation is very similar in the second case, where we have $78$ additional vector states. Here the additional $\mathbb{Z}_3$ involution
forces the $T^6$ lattice to have an E$_6$ structure in the geometric regime which becomes fully gauged at this locus. The precise form of the gauge enhancement can be made very precise by calculating the corresponding root lattice of the adjoint representation. In Section~\ref{sec:TargetSpace} we give explicit examples for those computations.
\item $\boldsymbol{\chi = \pm \frac12} \mathbf{(h^{1,1}+h^{2,1})}$ {\bf models :} The boundary of the classification in Figure \ref{fig:ScanPlot} is given by the lines with $ h^{2,1}=0$ or $h^{1,1}=0$. Models on the $h^{2,1}=0$ line correspond to {\it rigid} geometries i.e. Calabi-Yaus with no complex structure deformation in the NLSM interpretation. On the other hand we find models that have $h^{1,1}=0$. These are the mirror duals of those rigid geometries that have a fixed volume at the LGO locus. Those geometries cannot be obtained by the methods of toric geometry as a hypersurface or complete intersection in a given ambient space as those models necessarily have at least one K\"{a}hler modulus inherrited by the ambient space. Many models with a positive Euler number can be interpreted as the Landau-Ginzburg locus of toroidal orbifold compactifications. Examples of them are the $\mathbb{Z}_3$ or $\mathbb{Z}_3 \times \mathbb{Z}_3$ and their Landau-Ginzburg mirror that have gained a lot of attention in the past \cite{Dixon:1985jw,Aspinwall:2012qca,Chun:1989se}.
\item {\bf Generic models :} Here we have models with various distributions of $\mathbf{27}$ and $\overline{\mathbf{27}}$'s. Comparing the amount of E$_6$ charged fields with the Hodge numbers found in the classification of symmetric orbifold geometries \cite{Fischer:2012qj} lets us conclude that all of these models admit an orbifold phase.
\end{enumerate}
Finally we want to comment on the astonishing empirical relation satisfied by all the models of our classification that is:
\begin{align}
\label{eq:relation}
N_{\text{Add-S}}-3N_{\text{Add-V}}=(4-\mathcal{N})\cdot 76 \, ,
\end{align}
with $N_{\text{Add-S}}$ and $N_{\text{Add-V}}$ being the number of additional $\mathcal{N}=1$ chiral- and vector superfields neutral under the $\rm{E}_6 \times \rm{E}_8$  at the Fermat locus. This relation relates models with various matter content with their amount of supersymmetries.
 Note however that we count multiplets in $\mathcal{N}=1$ language even for higher amounts of supersymmetries. Hence in those cases it is more helpful to recollect the states in the higher SUSY representations. For example
the $\mathcal{N}=4$ vectormultiplet consists of one vector and three left chiral $\mathcal{N}=1$ fields. Hence in those cases the left and right hand side of the relation vanishes identically.
 However, 
the origin for $\mathcal{N}<4$ cases is unclear but might be explained by the strong symmetries and their anomalies at the Fermat locus. Moreover this relation might be a hint that all $A^9_1$ models might live in a common moduli as they have all been obtained from the same master models although they
have different amounts of supersymmetries. Finally we remark that many deformations away from the Fermat point leave this relation invariant. In all those Higgs transitions one vector-multiplet is traded for three chiral singlet fields.
\section{Tracing Target Space Symmetries}
\label{sec:TargetSpace}
In this section we want to have a closer look at two Landau-Ginzburg models and investigate their target space properties. Similar as in \cite{DISTLER199413,Distler:1994hq,Aspinwall:2011us} we investigate the deformation of the Landau Ginzburg theory and match those effects with the Higgs mechanism in the four dimensional theory. By using mirror symmetry those deformations can be interpreted as going to finite volume geometries and hence we match the spectrum with the ones from orbifold constructions.
\\
 In order for this to work we use the fact that Green-Plesser mirror symmetry actually works for {\it families} of mirror dual
CFT's which we review in the following. For this consider a CFT $\mathcal{C}$ where we act with the mirror map $\Gamma$ that acts as a left-$U(1)_R$ sign flip.
 However we can also deform the CFT $\mathcal{C}$ with an operator $\mathcal{U}$. Then it is always possible to define a mirror operator by the composition $\widetilde{\mathcal{U}}^{-1} \otimes \Gamma \otimes \mathcal{U}$ with $\widetilde{\mathcal{U}}^{-1}$ being the inverse deformation of $\mathcal{U}$ with the sign flip taken into account. This establishes the string geometries not only to be mirror dual at a specific point but over the whole moduli space summarized in the diagram of Figure \ref{fig:mirrorpairs}.
\begin{figure}[h!]
\begin{center}
\begin{tabular}{lcc}
$\Gamma(\mathcal{C})=\mathcal{C}_M$ & $\longleftrightarrow$ & $\widetilde{\mathcal{U}}(\mathcal{C}_M)$ \\
$\left\updownarrow\rule{0cm}{0.5cm}\right.  \renewcommand{\arraystretch}{0.5} \begin{array}{c}  ${\scriptsize Mirror}$  \\  ${\scriptsize map:}$ \, {\scriptsize \Gamma} \end{array} $&  & $\left\updownarrow\rule{0cm}{0.5cm}\right.$ \\
$\mathcal{C}$ & $\longleftrightarrow$ & $\mathcal{U}(\mathcal{C})$
\end{tabular}
\caption{\label{fig:mirrorpairs} Schematic graphic that shows how mirror symmetry at one point in moduli space extends to a whole set of (marginal) deformed CFT's.}
\end{center}
\end{figure}
This means for example when we know the CFT for a given Calabi-Yau X we can deform that CFT to a point where we know the mirror map, i.e. the Landau-Ginzburg point and its Fermat locus. There the mirror is given by the Green-Plesser map \eqref{eq:mirrormap}. By performing the inverse deformations  we obtain the mirror dual geometry to the former Calabi-Yau $\widetilde{X}$.\\
In our case we want to consider deformations away from the Landau-Ginzburg point towards large volume i.e.\ orbifold points. In the mirror these deformations are 
exactly polynomial deformations i.e. {\it complex structure} deformations away from the Fermat locus. Via this construction we can give an alternative description of large volume Calabi-Yau spaces as the mirror dual of a Landau-Ginzburg model where we can apply the techniques to compute spectra and symmetries.\footnote{In the smooth cases this is nothing but the Landau-Ginzburg-Calabi Yau correspondence \cite{Vafa:1988uu} as well as dual to orbifold models\cite{Chun:1989se}.} This strategy however has the limitation, that we trade the control of the K\"{a}hler moduli with those over the complex structure. \\\\
With this strategy at hand we want to investigate the symmetries of orbifold compactifications from the Landau-Ginzburg point. However these methods also apply for 
deformations to fully smooth phases or phases with no intermediate orbifold phase such as the Quintic. Moreover it is a particular benefit of this description that we can fully describe the low energy symmetries of non-factorisable orbifold geometries. \\
Calculations on non-factorisable orbifold tori are genuinely harder to perform  as the orbifold action does not respect holomorphicity of the target space coordinates \cite{Bizet:2013wha} whereas our methods are independent from that.\\\\
In the following we give two examples where we explicitly construct the full spectrum at the Fermat LGO point and compute the charges under all symmetries. It is worth noting that at those points we do not have any uncharged fields and hence no moduli dependence of superpotential couplings. We then perform the deformation to the orbifold
point in the mirror geometry which we match to the Higgs mechanism in the four dimensional theory. 
\subsection{The $SU(3)^4$ model}
\begin{figure}[t!]
\begin{picture}(0,140)(0,0)
\put(100,0){\includegraphics[scale=1]{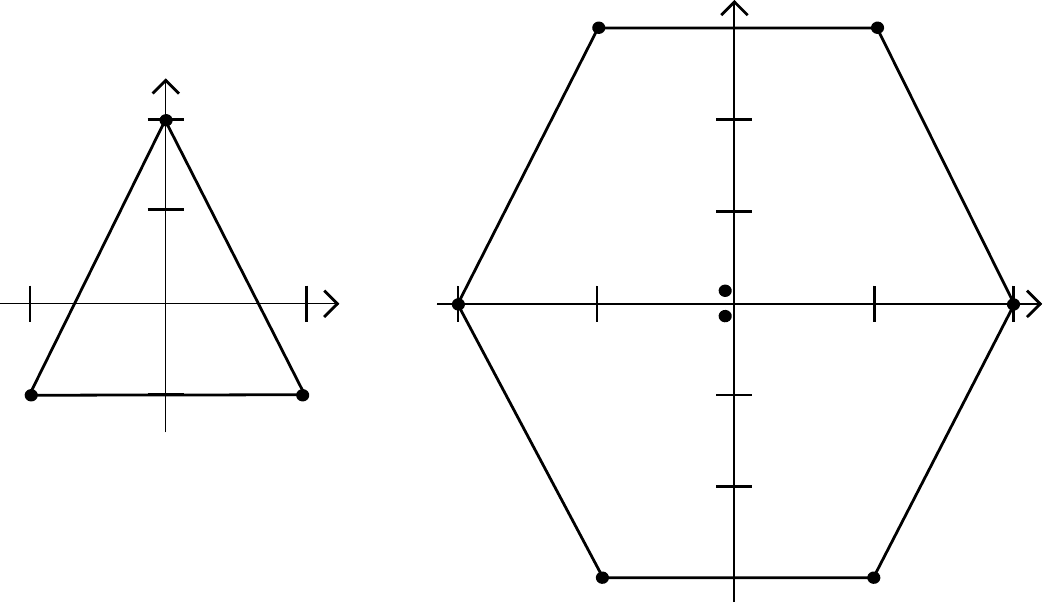}}
\put(200, 85){$q^B_{\text{iso}}$}
\put(155, 160){$q^B_{\text{str}}$}
\put(160, 140){{\tiny $|3;0,1 \rangle$ }}
\put(160, 131){{\tiny $|3;0,2\rangle$ }}

\put(195, 51){{\tiny $|5;0,1\rangle $}}
\put(195, 60){{\tiny$|1;0,2\rangle $}}

\put(70, 51){{\tiny $|1;0,1\rangle $}}
\put(70, 60){{\tiny$|5;0,2\rangle $}}

\put(225, 175){{\tiny  $\prod_a \overline{\phi}^{a,1}_{-\frac16} \overline{\psi}^{a,3}_0|5;0,1\rangle $}}
\put(350, 175){{\tiny$\prod_a \overline{\phi}^{a,1}_{-\frac16} \overline{\psi}^{a,2}_0|5;0,2\rangle $}}
\put(225, -2){{\tiny  $\prod_a \overline{\phi}^{a,2}_{-\frac16} \overline{\psi}^{a,1}_0|3;0,1\rangle $}}
\put(350, -2){{\tiny$\prod_a \overline{\phi}^{a,3}_{-\frac16} \overline{\psi}^{a,1}_0|3;0,2\rangle $}}
\put(240, 75){{\tiny  $\prod_a \overline{\phi}^{a,3}_{\frac16} \overline{\psi}^{a,2}_0|1;0,2\rangle $}}
\put(400, 75){{\tiny$\prod_a \overline{\phi}^{a,1}_{-\frac16} \overline{\psi}^{a,3}_0|1;0,1\rangle $}}
\put(315, 100){{\tiny  $\ldots|1;0,0\rangle $}}
\put(315, 90){{\tiny$\ldots|1;0,0\rangle $}}
\end{picture}
\caption{\label{fig:vaccharges}Charges of Vacua that form the weight lattice of the fundamental of $SU(3)^B$. On the right side we give the charges of 
the eight SU(3) gauginos under \eqref{eq:chargeB} resulting in the root lattice.}
\end{figure}
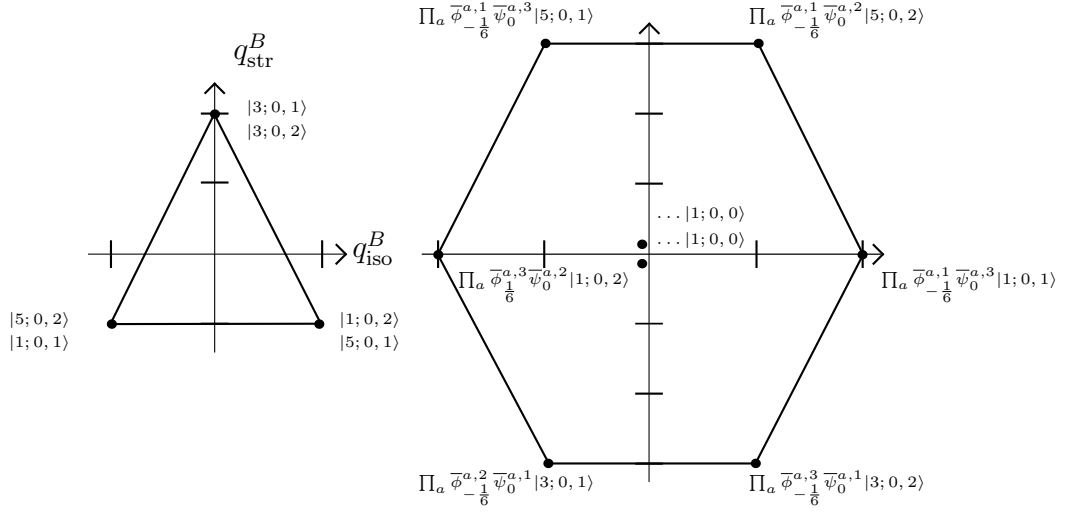

As a first example we consider a model with very peculiar symmetries in the LGO phase that we interpolate to the well known $\mathbb{Z}_3$ orbifold in deformed dual geometry. The model is described by the charge assignment given in Table \ref{tab:E6SU3Model} which differs from the one given in Appendix \ref{app:LGOList} by a charge redefinition for convenience. The GLSM of the dual geometry was considered in \cite{Blaszczyk:2011hs} which admits a $\mathbb{Z}_3$ orbifold phase with a non-factorized E$_6$ torus lattice and can be found in Table \ref{tab:SU(3)4GLSM} of Appendix \ref{app:SU(3)4GLSM}.
The charges of the mirror LGO can be obtained by applying the mirror map directly 
to the GLSM in Table \ref{tab:E6SU3Model}  and changing the charges for discrete ones.
\begin{table}[h!]
\begin{center}
\begin{tabular}{c| c c c| c c c| c c c|}
 & $\Phi^{1,1}$ & $\Phi^{1,2}$ & $\Phi^{1,3}$ & $\Phi^{2,1}$ & $\Phi^{2,2}$ & $\Phi^{2,3}$ & $\Phi^{3,1}$ & $\Phi^{3,2}$  & $\Phi^{3,3}$ \\ \hline  
$\rm{U}(1)_R$ & 1 & 1 & 1 & 1 & 1 & 1 & 1 & 1 & 1 \\ \hline
$\mathbb{Z}^{(1)}_3$ & 1 & 1 & 1 & -1 & -1 & -1 & 0 &0 & 0 \\
$\mathbb{Z}^{(2)}_3$ & 0 & 1 & -1 & 0 & 1 & -1 & 0 &1 & -1 \\ \hline
\end{tabular}
\caption{\label{tab:E6SU3Model}The charge assignment of the $\rm{SU}(3)^4$ mirror LGO. }
\end{center}
\end{table}
Due to the symmetric structure of the $\mathbb{Z}_3$ charges it is convenient to sub-label the 9 fields according to $i= 3\cdot a + j$ with $a,j = 1,2,3$. We use the same labeling in the geometric GLSM where the $a$ index identifies a torus an interpretation which is lost in the mirror dual LGO.
The spectrum is summarized in the following as
\begin{center}
\begin{tabular}{|c|cccc|}\hline
&Singlets & $\mathbf{27}$ & $\overline{\mathbf{27}}$ & $E_6$-Adjoint \\ \hline
Left-Chirals& 324 & 36  & 0 & 0 \\
Vectors& 32  & 0 & 0 & 1 \\\hline
\end{tabular}
\end{center}
The 32 vector multiplets assemble themselves into four adjoints of an additional $SU(3)^4$ gauge group and the 324 singlets form tri-fundamental representations under those. The full matter content and all its discrete charges are given at the end of this subsection in Table \ref{tab:27plets} and \ref{tab:singlets}.
The explicit form of the 32 vector multiplets with respect to the oscillator states can be found in Appendix \ref{tab:SU34Vectors}. Note that unlike as in the
$\mathcal{N}=4$ model on the $E_6$ torus lattice not all winding modes survive the $\mathbb{Z}_3$ projection. However the $\mathbb{Z}_3$ projection can be interpreted as an adjoint breaking of E$_6$ to $SU(3)^3$'s. Besides the three geometrical interpreted $SU(3)$'s the fourth one comes out of the E$_8$ and stays preserved. \\\\
In order to capture the quantum numbers of the four $SU(3)$ generators we collect their eight Cartan generators in four pairs
of 'strangeness and isospin' $(q^{X}_{\text{str}},q^X_{\text{iso}})$ where we label the four SU(3)'s by  letters $X \in \{ A,B,C, D \}$.
The SU(3) Cartan operators are the following sums in our two index notation:
\begin{align}
\label{eq:chargeA}
q^A_{\text{str}} = \sum^3_{i=1}\left( -2 q_{3,i}+q_{1,i}+q_{2,i}\, \right) , \\
q^A_{\text{iso}} = \sum^3_{i=1} \left( q_{1,j}-q_{2,j}\right)\, ,
\end{align}
while the other ones have the following index structure:
\begin{align}
\label{eq:chargeB}
q^B_{\text{str}} = \sum^3_{a=1}\left(-2 q_{a,1}+q_{a,2}+q_{a,3} \right)\, , \\
q^B_{\text{iso}} = \sum^3_{a=1}\left( q_{a,2}-q_{a,3}\right)\, . 
\end{align}
The structure is similar for the following two pairs of charge operators but to keep the
index summation structure similar, all indices are to be understood as $i,a >0$ mod three.
Then the Cartans of $SU(3)^C$ can be written as
\begin{align}
\label{eq:chargeC}
q^C_{\text{str}} = \sum^3_{a=1}\left( -2 q_{a,a}+q_{a,a+1}+q_{a,a+2}\right)\, , \\
q^C_{\text{iso}} = \sum^3_{a=1}\left( q_{a,a+1}-q_{a,a+2}\right)\, ,
\end{align}
and the final set of charge operators are given by
\begin{align}
\label{eq:su34generators}
q^D_{\text{str}} = \sum^3_{a=1}\left( -2 q_{a,1-a}+q_{a,2-a}+q_{a,-a}\right)\, , \\
q^D_{\text{iso}} = \sum^3_{a=1}\left( q_{a,-a}-q_{a,2-a}\right)\, .
\end{align}
In order to construct the R-symmetry charge operator we have to divide out 
the strangeness operator of the four $SU(3)$'s to obtain
\begin{align}
Q^{(18)}_R = 3k_0 - 2q_- + 2q^{(A)}_{\text{str}} + 2q^{(B)}_{\text{str}} +2q^{(C)}_{\text{str}} +2q^{(D)}_{\text{str}} \text{ mod }18 \, ,
\end{align}
which guarantees uniform charge of all charginos. In addition the two orbifold twists induce
two discrete symmetries with charge generators
\begin{align}
Q^1_{\mathbb{Z}_3} =& k_1 + q^A_{\text{iso}} + q^C_{\text{iso}} + q^D_{\text{iso}} \text { mod } 3\, , \\
Q^2_{\mathbb{Z}_3} =& k_2 + q^B_{\text{iso}} + q^C_{\text{iso}} - q^D_{\text{iso}} \text { mod } 3\, ,
\end{align}
corrected by the isospin Cartan operators.
Calculating strangeness and isospin quantum numbers of vacua in other twisted sectors we find that they always form fundamental representations under the four SU(3)'s.  For $SU(3)^B$ we depict non-trivial charged vacua and adjoint gauginos in Figure \ref{fig:vaccharges}.
Note that all gauginos have R-charge 3 under the 4D R-symmetry. However before we come to the 4D superpotential we first construct the
charges of all left-chiral superfields.\\ The 36 $\mathbf{27}$-plets are summarized in Table \ref{tab:27plets}.
\begin{table}[t!]
\begin{center}
\begin{tabular}{|c|ccc|c|}\hline
label&$\rm{E}_6 \times \rm{SU}(3)^4$ Repr. & $Q^1_{\mathbb{Z}_3}$ & $Q^2_{\mathbb{Z}_3} $ & $Q_R$ Superfield \\ \hline
$\mathbf{27}_{A^1}$&$(\mathbf{27},\mathbf{3},\mathbf{1},\mathbf{1},\mathbf{1})$ & 2 & 0  & 6 \\ 
$\mathbf{27}_{A^2}$&$(\mathbf{27},\mathbf{3},\mathbf{1},\mathbf{1},\mathbf{1})$ & 1 & 0  & 6 \\ 
$\mathbf{27}_{A^3}$&$(\mathbf{27},\mathbf{3},\mathbf{1},\mathbf{1},\mathbf{1})$ & 0 & 0  & 0\\ \hline
$\mathbf{27}_{B^1}$&$(\mathbf{27},\mathbf{1},\mathbf{3},\mathbf{1},\mathbf{1})$ & 0 & 1 & 6 \\ 
$\mathbf{27}_{B^2}$&$(\mathbf{27},\mathbf{1},\mathbf{3},\mathbf{1},\mathbf{1})$ & 0 & 2 & 6\\ 
$\mathbf{27}_{B^3}$&$(\mathbf{27},\mathbf{1},\mathbf{3},\mathbf{1},\mathbf{1})$ & 0 & 0 &  0  \\ \hline
$\mathbf{27}_{C^1}$&$(\mathbf{27},\mathbf{1},\mathbf{1},\mathbf{3},\mathbf{1})$ & 2 & 2 & 6\\ 
$\mathbf{27}_{C^2}$&$(\mathbf{27},\mathbf{1},\mathbf{1},\mathbf{3},\mathbf{1})$ & 1 & 1 &  6\\ 
$\mathbf{27}_{C^3}$&$(\mathbf{27},\mathbf{1},\mathbf{1},\mathbf{3},\mathbf{1})$ & 0 & 0 &  0\\ \hline
$\mathbf{27}_{D^1}$&$(\mathbf{27},\mathbf{1},\mathbf{1},\mathbf{1},\mathbf{3})$ & 2 & 1 &  6 \\ 
$\mathbf{27}_{D^2}$&$(\mathbf{27},\mathbf{1},\mathbf{1},\mathbf{1},\mathbf{3})$ & 1 & 2 &  6\\ 
$\mathbf{27}_{D^3}$&$(\mathbf{27},\mathbf{1},\mathbf{1},\mathbf{1},\mathbf{3})$ & 0 & 0 &  0\\ \hline
\end{tabular}
\caption{\label{tab:27plets}All quantum numbers of the $\mathbf{27}$ representations of E$_6$}
\end{center}
\end{table}
We verify that indeed all states are distinguished by individual quantum numbers.
 We get a similar structure for the 324 additional $\rm{E}_6$ singlets. These fields form tri-fundamental representations of the $\rm{SU}(3)^4$ gauge factors summarized in Table \ref{tab:singlets}.
\begin{table}[h]
\begin{center}
\begin{tabular}{|c|ccc|c|}\hline
label & $\rm{E}_6 \times \rm{SU}(3)^4$ Repr. & $Q^1_{\mathbb{Z}_3}$ & $Q^2_{\mathbb{Z}_3} $ &  $Q_R$ \\ \hline
$S_{a_1}$&$(\mathbf{1},\mathbf{1},\overline{\mathbf{3}},\overline{\mathbf{3}},\overline{\mathbf{3}})$ & 1 & 0  & 6\\ 
$S_{a_2}$&$(\mathbf{1},\mathbf{1},\overline{\mathbf{3}},\overline{\mathbf{3}},\overline{\mathbf{3}})$ & 2 & 0  & 6\\ 
$S_{a_3}$&$(\mathbf{1},\mathbf{1},\overline{\mathbf{3}},\overline{\mathbf{3}},\overline{\mathbf{3}})$ &  0 & 0 & 0 \\ \hline
$S_{b_1}$&$(\mathbf{1},\overline{\mathbf{3}},\mathbf{1},\overline{\mathbf{3}},\overline{\mathbf{3}})$ & 0 & 1 &  6 \\ 
$S_{b_2}$&$(\mathbf{1},\overline{\mathbf{3}},\mathbf{1},\overline{\mathbf{3}},\overline{\mathbf{3}})$ & 0 & 2 & 6\\ 
$S_{b_3}$&$(\mathbf{1},\overline{\mathbf{3}},\mathbf{1},\overline{\mathbf{3}},\overline{\mathbf{3}})$ & 0 & 0 & 0 \\ \hline 
$S_{c_1}$&$(\mathbf{1},\overline{\mathbf{3}},\overline{\mathbf{3}},\mathbf{1},\overline{\mathbf{3}})$ & 1 & 1 & 6\\
$S_{c_2}$&$(\mathbf{1},\overline{\mathbf{3}},\overline{\mathbf{3}},\mathbf{1},\overline{\mathbf{3}})$ & 2 & 2 & 6 \\
$S_{c_3}$&$(\mathbf{1},\overline{\mathbf{3}},\overline{\mathbf{3}},\mathbf{1},\overline{\mathbf{3}})$  & 0 & 0 & 0\\ \hline
$S_{d_1}$&$(\mathbf{1},\overline{\mathbf{3}},\overline{\mathbf{3}},\overline{\mathbf{3}},\mathbf{1})$ & 1 & 2 &6 \\ 
$S_{d_2}$&$(\mathbf{1},\overline{\mathbf{3}},\overline{\mathbf{3}},\overline{\mathbf{3}},\mathbf{1})$ & 2 & 1 & 6\\ 
$S_{d_3}$&$(\mathbf{1},\overline{\mathbf{3}},\overline{\mathbf{3}},\overline{\mathbf{3}},\mathbf{1})$ & 0 & 0 & 0\\ \hline
\end{tabular}
\caption{The gauge representation of the 324  $\rm{E}_6$ singlet states and their R-charges.}
\label{tab:singlets}
 \end{center}
\end{table}
We note two important facts about the spectrum: First we do not find any uncharged fields i.e.
all fields have non-trivial charges under some operator and thus there is no modulus dependence of the 4D superpotential. Secondly we find that the spectrum is completely invariant under the permutation of the four SU(3) gauge factors combined with a reshuffling of discrete quantum numbers. We conclude that we actually have a $SU(3)^4 \rtimes  S_4$ gauge group. This $S_4$ permutation symmetry is simply a descendant of the permutation symmetry of the chiral world sheet fields on the LGO side. Indeed on the 2D theory this permutation is achieved by a charge redefinition that multiplies $\mathbb{Z}^{(1)}_3$ and $\mathbb{Z}^{(2)}_3$ charges
of the superfields in Table \ref{tab:E6SU3Model} by factors of two.
This permutation symmetry makes also the interpretation of the origin of the four SU(3) factors ambiguous as any of them could come out of the $E_8$ while the others ones could come from the geometry.\\\\
{\bf Landau-Ginzburg Deformation}\\\\
In the following we want to deform the Landau-Ginzburg superpotential away from the Fermat locus. This is possible by the following deformation terms 
that we can add to the Fermat superpotential
\begin{align}
 \mathcal{W}_{\text{Deform}} =A_a \Phi^{a,1}\Phi^{a,2}\Phi^{a,3} +  B_{i}\Phi^{1,i}\Phi^{2,i}\Phi^{3,i}  + C_i \Phi^{1,1+i}\Phi^{2,2+i}\Phi^{3,i} + D_i \Phi^{1,1+i}\Phi^{2,3+i}\Phi^{3,2+i}  \, ,
\end{align}
which exhaust all possible deformation terms in accord with the symmetries of the 2D theory.
In total we have 12 deformations that we split into $4 \times 3$ triples. Switching on any deformation, say $B_1$, results in the reduction of the spectrum given in Table~\ref{eq:SU(3)4LGOdefspectrum} which we have computed on the LGO side.
\begin{table}[t]
\begin{center}
 \begin{tabular}{|c|c|c|c|}
 \hline
$\begin{array}{c} \text{Amount of} \\ \text{deformation(s)} \end{array}$ & $\begin{array}{c} \text{non-Fermat} \\ \text{deformation(s)} \end{array}$
 & Chiral $E_{6}$ Singlets & Vector $E_{6}$ Singlets \\
  \hline
 0& - & 324 & 32 \\
  \hline
  1 & $B_{i}$ & 264 & 12 \\
  \hline
  2 & $B_{1},B_{2}$ & 258 & 10\\
  \hline
 3 & $B_{1},B_{2},B_{3}$ & 252 & 8 \\
  \hline
 \end{tabular}
\caption{\label{eq:SU(3)4LGOdefspectrum}$E_6$ uncharged spectra of the SU$(3)^4$ LGO for one kind of deformation. Changing $B_i$ any other deformation results in the same spectrum. The number of $\mathbf{27}$'s is deformation invariant. }
\end{center}
\end{table}
We observe that the single term, say $B_1 \neq 0$ results in the following breakdown of the four dimensional gauge group:
\begin{align}
 \rm{SU}(3)^A \times \rm{SU}(3)^B \times \rm{SU}(3)^C \times \rm{SU}(3)^D \xrightarrow{B_i \neq 0}\rm{SU}(3)^B \times \rm{U}(1)^4 \, .
\end{align}
Of course this breaking is exactly the same for another choice of $X_i \neq 0$ and leaves the $SU(3)^X$ gauge factor unbroken while breaking the three SU(3)'s to two $U(1)$'s.\footnote{The same breaking patter has been observed in \cite{Beye:2014nxa} on the CFT level for $T^2/\mathbb{Z}_3$ orbifolds when moving away from from the self-dual radius. However note that we do not have a factorized geometry.} The equivalence of the deformation parameters and their induced breaking is again a result of the permutation symmetry of the WS fields.\\
Adding additional $X_j$ deformations  of the same kind leaves the $SU(3)^X$ unbroken but breaks the remnant U(1) factors completely.\\
However switching on any deformation of two different kinds, say $A_i, B_j \neq 0$ breaks all SU(3) gauge factors.\\\\
 The deformations $X_i$ have an interpretation in form of torus K\"{a}hler moduli in the mirror geometry. 
 Indeed we find the spectrum we expect from the $\mathbb{Z}_3$ orbifold \cite{DIXON1985678} given by
\begin{align}
\label{eq:OrbifoldSpectrum}
\begin{array}{rrll}
\text{ Untwisted Sector :}& 3 \times (\mathbf{27},\mathbf{3}) \, &\text{K\"{a}hler moduli:} &9 \times (\mathbf{1},\mathbf{1}) \, , \\
\text{27 Fixed Points :} &   (\mathbf{27},\mathbf{1}) &\text{Bundle moduli:} &3 \times (\mathbf{1},\overline{\mathbf{3}}) \, .
\end{array}
\end{align}
which coincides with our computation on the LGO side given in the last column of Table~\ref{eq:SU(3)4LGOdefspectrum}. 
Adding additional deformations corresponds to blow/ups of the orbifold singularities.
In the following we match the 2D LGO deformations with the Higgs mechanism in the four dimensional theory. \\\\
{\bf Match with 4D effective action}\\\\
As we have all symmetries at the LGO Fermat point we can write down the full 4D effective superpotential up to trilinear order and match the 2D LGO deformation with
the effects of the four dimensional Higgs mechanism. Before we do that we consider the four dimensional superpotential $\mathcal{W}_{4D}$ given at tree-level at the LGO Fermat locus 
\begin{align}
\label{eq:W4DSU34}
\mathcal{W}_{4D} \ni& \sum_I^{A,B,C,D} \left|\epsilon_{ijk}\right|\mathbf{27}_{I^i}\mathbf{27}_{I^j}\mathbf{27}_{I^k} + \sum_I^{a,b,c,d} \left|\epsilon_{ijk}\right| S_{I_i}S_{I_j}S_{I_k} \, .
\end{align}
The structure of all tree-level couplings is very compact due to the large amount of symmetries of the model. Note that there are no cubic couplings of the fields with themselves. This is guaranteed by the 4D R-symmetry of the superpotential $Q^{(18)}_R(\mathcal{W}) = -6$ mod 18. Couplings with the singlets $S_{I_i}$ and $\mathbf{27}$-plets only occur at the fourth order.
The deformations of the LGO $A,B,C,D \neq 0$ can be matched to be non-vanishing VEVs inside tri-fundamentals $S_{A_3}, S_{B_3}, S_{C_3}$ or $S_{D_3}$ of Table \ref{tab:singlets}. We focus again w.l.o.g.\ on
$B_i$ deformation that gives a VEV in the 
$(\mathbf{1},\overline{\mathbf{3}},\mathbf{1},\overline{\mathbf{3}},\overline{\mathbf{3}})_{b_3}$ representation. \\ 
In the following we want to discuss the D-and F-flat directions
and the mass-terms explicitly. To do so, we need to introduce the three indices
 $S_{b_3} \rightarrow S_{b_3}^{i,j,k}$ with $i,j,k$ being anti-fundamentals indices of $\rm{SU}(3)^A,\rm{SU}(3)^C$ and $\rm{SU}(3)^D$, respectively.
The VEVs $ b_i \leftrightarrow B_i$ lie in the fields as
\begin{align}
 B_1:& \langle S_{b_3}^{1,1,1}\rangle  = \langle S_{b_3}^{2,2,2}\rangle =\langle S_{b_3}^{3,3,3}\rangle =  b_1  \, , \\ 
 B_2:& \langle S_{b_3}^{1,2,3}\rangle  = \langle S_{b_3}^{2,3,1}\rangle =\langle S_{b_3}^{3,1,2}\rangle =  b_2  \, , \\ 
 B_3:& \langle S_{b_3}^{2,1,3}\rangle  = \langle S_{b_3}^{1,3,2}\rangle =\langle S_{b_3}^{3,2,1}\rangle = b_3  \, .
\end{align}
which guarantees D-flatness. F-flatness is checked by noting in \eqref{eq:W4DSU34} that there is no quadratic $S_{b_3}$ coupling but always with fields
that acquire no VEV.
The corresponding couplings of the above fields in \eqref{eq:W4DSU34} give the mass terms after inserting the VEVs:
\begin{align}
\label{eq:Z3Massmatrix}
\mathcal{W}_{4D} &\ni  \epsilon_{xyz}\epsilon_{ilo}\epsilon_{jmp} \epsilon_{knq} S_{b_x}^{i,j,k} S_{b_y}^{l,m,n} S_{b_z}^{o,p,q} \,.
\end{align}
The indices $i,j,k,l,m,n,o,p,q$ are the three SU(3) indices, while $x,y,z$ label the three $S_{b_x}$ states. The rank of the mass matrix for the 1,2 and three deformations, the Goldstinos and the amount of massive chiral superfields is summarized in the following table:
\begin{center}
 \begin{tabular}{|c|c|c|c|}
 \hline
$\begin{array}{c} \text{Field Theory} \\ \text{VEVs} \end{array}$ & $\begin{array}{c} \text{Mass Matrix} \\ \text{Rank} \end{array}$ & $\begin{array}{c} \text{Goldstone} \\ \text{Bosons} \end{array}$ & $\begin{array}{c} \text{Massive} \\ \text{Superfields} \end{array}$ \\
  \hline
  - & 0 & 0 & 0\\
  \hline
  $b_{1}$ & 20 & 20 & $2\cdot20+20$ \\
  \hline
  $b_{1},b_{2}$ & 22 & 22  & $2\cdot22+22$ \\
  \hline
  $b_{1},b_{2},b_{3}$ & 24 & 24 & $2\cdot 24 + 24$ \\
  \hline
 \end{tabular}
\end{center}
We find that the missing multiplets exactly accommodate for the change in the singlet numbers from the LGO calculation given in Table~\ref{eq:SU(3)4LGOdefspectrum}.\\
At next we can turn towards the $\mathbf{27}$ fields and their couplings. Upon the breakdown of the fields the nine $\mathbf{27}_{A^i},\mathbf{27}_{C^i}$ and $\mathbf{27}_{D^i}$ fields decompose in
the 27 states that we interpret as the twisted fields at the fixed points in the orbifold language. Furthermore, the following tree level couplings among $\mathbf{27}$-plets are generated by singlet VEV insertions of fourth order:
\begin{align}
\begin{split}
\mathcal{W}_{\rm 4D-Higgsed} \in& \sum_i^3 \mathbf{27}_{A^i} \mathbf{27}_{C^i} \mathbf{27}_{D^i} S_{b_i} \, \\
=& b_1(\sum^3_m\mathbf{27}^{m,m,m}_{A^1} \mathbf{27}^{m,m,m}_{C^1} \mathbf{27}^{m,m,m}_{D^1}) \\
+& b_2(\sum^3_m\mathbf{27}^{m, m+1,m+2}_{A^2} \mathbf{27}^{m, m+1,m+2}_{C^2} \mathbf{27}^{m, m+1,m+2}_{D^2}) \\
+& b_3(\sum^3_m\mathbf{27}^{m, m+2,m+1}_{A^3} \mathbf{27}^{m, m+2,m+1}_{C^3} \mathbf{27}^{m, m+2,m+1}_{D^3}) \, ,
\end{split}
\end{align} 
which we interpret as couplings among the various fields located at different fixed points in the orbifold language whereas the VEVs play the role of the world sheet instanton effects that communicate the coupling. Note that it is straightforward from this perspective to obtain all order couplings and their moduli dependence just by inserting the VEVs in the superpotential obtained as the Fermat point. Finally note, that the deformation to the orbifold phase does not break the R-symmetry of the four dimensional theory as expected.\\\\
At this point we have not exhausted all Landau Ginzburg deformations yet.
Indeed there are nine residual ones $A_i,C_i$ and $D_i$ corresponding to blow-up modes. However in the general orbifold there are 27 independent blow ups
and hence we have a LGO description where three orbifold fixed points get blown up simultaneously. It is remarkable that all deformations correspond to Higgs VEVs in 4D fields that are neutral under the R-symmetry. Hence we can resolve all 27 singularities while keeping the R-symmetry. This is in contrast to the usual expectation that there are no R-symmetries in smooth heterotic compactifications. However note that we have a non-factorized E$_6$ root lattice underlying the orbifold geometry and that a single LGO deformation always blows-up three fixed points with the same volume. Hence along this very specific direction in the K\"{a}hler moduli space the R-symmetry is preserved as similarly observed in \cite{Ludeling:2012cu}. Similarly observations have been made in smooth compactifications \cite{Anderson:2009nt} where U(1) become massless along special directions in the K"{a}hler moduli space. \\
Also note that in this description all K\"{a}hler moduli are treated completely democratic thanks to the $S_4$ permutation symmetry. Hence in the smooth case we can not distinguish blow-up from ambient volume.

\subsection{The $\mathbb{Z}_3 \times \mathbb{Z}_{3,\text{free}}$ Mirror LGO}
As a second example we consider the $\mathbb{Z}_3$ orbifold and divide by an additional $\mathbb{Z}_{3,\text{free}}$ action and construct its mirror dual
Landau-Ginzburg orbifold. The Table~\ref{tab:RotoGLSM} of Appendix~\ref{app:rotoglsm}. But before we consider the Landau-Ginzburg model we construct the $\mathbb{Z}_3 \times \mathbb{Z}_{3,\text{free}}$ orbifold spectrum
that we want to match.\\\\
{\bf Spectrum of the  $\mathbf{\mathbb{Z}_3 \times \mathbb{Z}_{3,\text{free}}}$ orbifold }\\\\
The orbifold model is described by the geometric twist vectors $v^i$ and gauge embedding $V^i$ that are modded out of the $T^6$ by their exponential action and have the form
\begin{align}
v^i_3 =& (\frac13,\frac13,-\frac23)\, , \text{ and } V^i_3 = (v^i_3,0^5)(0^8) \, , \\
v^i_{\text{3,free}} =& (\frac13,-\frac13,0)\, , \text{  and } V^i_{\text{3,free}} = (v^i_{\text{3,free}},0^6)(0^8) \, ,
 \end{align}
where the second twist vector $v^i_{\text{3,free}}$ is combined with a lattice shift such that it acts freely on $T^6/\mathbb{Z}_3$ and with its gauge shift embedding $V^i_{\text{3,free}}$
inside the first $E_8$ gauge factor. The additional roto-translational embedding does not induce additional fixed points but acts 
as an additional projection in the untwisted sector and a triple-wise identification of fixed points. In the following we use represent a state as $|q\rangle_R \otimes | P\rangle_L$ with $E_8$ weight vector $P$ and SO(8) weight $q$ using the conventions in  \cite{Vaudrevange:2008sm}. Also recall the projection condition in the untwisted sector \begin{align}
e^{i2 \pi i (P\cdot V + q \cdot v)  } = \text{id} \, .
\end{align}
 Hence starting from the orbifold spectrum in \ref{eq:OrbifoldSpectrum} the untwisted spectrum is projected in the following way: K\"{a}hler moduli of the form
\begin{align}
\label{eq:Kaehler}
| 0, \underline{1,0,0} \rangle_R \otimes \alpha^{\overline{i}}|0 \rangle_L 
\end{align}
are reduced from nine to only three diagonal ones.
The six roots of the SU(3) gauge symmetry that have the explicit form
\begin{align}
\alpha^\mu |0 \rangle_R \otimes |  (\underline{1,-1,0}),0^5)(0^8) \rangle_L \, ,
\end{align}
get all projected out by the freely acting twist such that only their two Cartan generators survive. Similarly the the three $(\mathbf{27},\mathbf{3})$-plets in the untwisted sector get reduced to only the three (bosonic) $\mathbf{27}$'s
\begin{center}
{
\renewcommand{\tabcolsep}{0.01cm}
\begin{tabular}{lll}
$|0, 0 ,0, 1 \rangle_R \otimes$  $\left\{\renewcommand{\arraystretch}{1.5} \begin{array}{l} |(0,0,1,\pm(\underline{ 1,0^4}))(0^8) \rangle_L \\
|(-1,-1,0^5)(0^8)\rangle_L \\
|(-\frac12,-\frac12,\frac12,(\pm \frac12)^5) \rangle_L
    \end{array}  \right.$\ \ \ &

 $|0, 1 , 0 ,0 \rangle_R \otimes$  $|\left\{\renewcommand{\arraystretch}{1.5} \begin{array}{l} (0,-1,0,\pm(\underline{ 1,0^4}))(0^8) \rangle_L\\
|(1,0,10^5)(0^8)\rangle_L \\
|(\frac12, -\frac12, \frac12,(\pm \frac12)^5)\rangle_L
    \end{array}  \right.$ \\ 
		
\end{tabular}}
\end{center}
\begin{center}
{
\begin{tabular}{l}
 $|0, 0 , 1 , 0 \rangle_L \otimes$  $\left\{\renewcommand{\arraystretch}{1.5} \begin{array}{l} |(-1,0,0,\pm(\underline{ 1,0^4}))(0^8)\rangle_L \\
|(0,1,10^5)(0^8)\rangle_L \\
|(-\frac12, \frac12, \frac12,(\pm \frac12)^5)\rangle_L
    \end{array}  \right.$ 
\end{tabular}
}
\end{center}
whose roots we have split up in the ($2\cdot 5 + 1 + 16$) contributions.
The free $\mathbb{Z}_3$ involution identifies three fixed points but acts trivial on the spectrum. Hence in total
we have the following spectrum.
\begin{align}
\label{eq:OrbifoldZ3Spectrum}
\begin{array}{rrll}
\text{ Untwisted Sector : }& 3 \times \mathbf{27} \, &\text{K\"{a}hler moduli} & 3 \times \mathbf{1} \, , \\
\text{9 Fixed Points : } &   \mathbf{27} &\text{Bundle moduli} & 9 \times \mathbf{1} \, .
\end{array}
\end{align}
The GLSM for the above geometry can be realized by the charge assignment given in Table \ref{tab:RotoGLSM} of Appendix \ref{app:rotoglsm} which is simply the GLSM that admits the $\mathbb{Z}_3$ orbifold phase supplemented with another $\mathbb{Z}_3$ element.\\\\
{\bf The $\mathbb{Z}_3 \times \mathbb{Z}_{3,\text{free}}$ Landau-Ginzburg model}\\\\
We construct the mirror-dual Landau-Ginzburg orbifold by taking the orthogonal $\mathbb{Z}_3$ charge assignment of the GLSM in Table \ref{tab:RotoGLSM} and obtain the mirror LGO with charges given in Table~\ref{tab:RotoLgor}.
\begin{table}
\begin{center}
\begin{tabular}{c| c c c| c c c| c c c|}
 & $\Phi^{1,1}$ & $\Phi^{1,2}$ & $\Phi^{1,3}$ & $\Phi^{2,1}$ & $\Phi^{2,2}$ & $\Phi^{2,3}$ & $\Phi^{3,1}$  & $\Phi^{3,2}$ & $\Phi^{3,3}$ \\ \hline  
$\rm{U}(1)_R$ & 1 & 1 & 1 & 1 & 1 & 1 & 1 & 1 & 1 \\ \hline
$\mathbb{Z}^1_3$ & 1 & 0 & -1 & 0 & 0 & 0 & -1 & 1 & 0 \\
$\mathbb{Z}^2_3$  & 0 & 1 & -1 & 0 & 0 & 0 & 1 & 1 & 1 \\ 
$\mathbb{Z}^3_3$  & 0 & 0 & 0 & 1 & -1 & 0 & 1 & 1 & 1 \\ \hline
\end{tabular}
\caption{\label{tab:RotoLgor}The charge assignment of the $\mathbb{Z}_3 \times \mathbb{Z}_{3,\text{free}}$ Mirror-LGO.}
\end{center}
\end{table}
Computing the full massless spectrum at the Fermat locus gives the following spectrum:
\begin{center}
\begin{tabular}{|c|cccc|}\hline
Repr & Non $E_6$ & $\mathbf{27}$ & $\overline{\mathbf{27}}$ & $E_6$ Adjoints \\ \hline
Left-Chirals & 252 & 12 & 0 & 0\\ 
Vectors & 8 & 0 & 0 & 1\\ \hline
\end{tabular}
\end{center}
We find the eight Landau-Ginzburg U(1)'s and additional singlet fields. This time we do not have have a non-Abelian enhancement and hence we can not give the spectrum in the same compact way as before. We obtain the simplest 4D R-charge generator as
\begin{align}
\label{eq:RotoRcharge}
Q^{(18)}_R = 3k_0 - 2q_- \text{ mod } 18 \, ,
\end{align}
under which the superpotential is charged $q_R(W_{4D})=-6\text{ mod }18$.
For the U(1) generators we simply stick with the original charge formulas given in \eqref{eq:Z37model} which results in fractional charges.
In the following we focus on the Higgs fields and the fields that become massive when we deform to the orbifold phase in the mirror.
\\\\
{\bf Landau-Ginzburg Deformation}\\\\
At next we look at the LGO deformations away from the Fermat point. They are given by the following four deformations 
\begin{align}
\mathcal{W}_{\text{Deform}} = \sum^3_{a=1} A_a  \Phi^{a,1}\Phi^{a,2} \Phi^{a,3}  + B \Phi^{1,3} \Phi^{2,2}\Phi^{3,2}
\end{align}
We can compute the resulting spectrum upon switching on the various deformation terms summarized in Table \ref{eq:SU(3)4LGOspectrum}.
\begin{table}[t]
\begin{center}
 \begin{tabular}{|c|c|c|c|}
 \hline
$\begin{array}{c} \text{Amount of} \\ \text{deformation(s)} \end{array}$ & $\begin{array}{c} \text{non-Fermat} \\ \text{deformation(s)} \end{array}$
 & Chiral $E_{6}$ Singlets & Vector $E_{6}$ Singlets \\
  \hline
 0& - & 252 & 8 \\
  \hline
  1 & $A_{a}/B $ & 196 & 6 \\
  \hline
  2 & $A_{1},A_{2}$ & 140 & 4\\
  \hline
 3 & $A_{1},A_{2},A_{3}$ & 84 & 2 \\
  \hline
 \end{tabular}
\caption{\label{eq:SU(3)4LGOspectrum} Spectra of the SU$(3)^4$ LGO for one kind of deformation. Changing $A_a$ any other deformation results in the same spectrum. }
\end{center}
\end{table}
Any deformation removes two vectors and 56 singlet fields from the spectrum. By comparing the spectrum we find that
switching on all three $A_a$ deformations brings us to the orbifold phase where the massless spectrum matches the one on the CFT level
\eqref{eq:OrbifoldZ3Spectrum} perfectly. Hence the three orbifold K\"{a}hler moduli correspond to the deformations $A_a$. Thus again we interpret the LGO deformations $A_a$ as giving the three tori finite size.
The leftover deformation $B$ is then the blow-up deformation of all nine fixed points simultaneously.
\\\\
{\bf Match with 4D effective action}\\\\
Again we match the deformation of the Fermat LGO with the Higgs mechanism in the 4D theory. As the spectrum is much less compact as in the $SU(3)^4$ we omit showing it here in all details but consider the Higgs fields in more detail. The $(1;0,0,0)$ twisted sector has twelve neutral fermions that can be collected to four groups corresponding to the respective LGO deformation. We give the states in the following notation
\begin{align}
 S_{a,b,c;i,j,k}: \quad \phi^{a,i}_{-\frac56} \phi^{b,j}_{-\frac56} \psi^{c,k}_{\frac23}|1;0,0,0\rangle \, .
\end{align}
 The three Higgs states that obtain a VEV corresponding to $A_a$ deformations
are given by the $a=b=c$ which we can, in the GLSM analogy, identify as the ambient torus deformations:
\begin{align}
 S_{a,a,a;1,2,3} \, , \quad  S_{a,a,a;1,3,2} \, , \quad  S_{a,a,a;2,3,1} \, .
\end{align}
These states have non-trivial charges under $\left(U(1)^{a,1},U(1)^{a,2},U(1)^{a,3}\right)$ as
\begin{align}
\left(q_{a,1},q_{a,2},q_{a,3}\right)(S_{b,b,b;\underline{1,2,3}}) =\delta_{a,b} \left(\underline{1/3,1/3,-2/3}\right) \, ,
\end{align}
where we have highlighted the permutations of the indices and charges.
The last three states can be identified with the Higgs-fields of the $B$ deformations and are given as
\begin{align}
 S_{1,2,3;3,2,2} \, , \quad  S_{1,3,2;3,2,2} \, , \quad  S_{2,3,1;2,2,3} \, .
\end{align}
 The Higgs fields that correspond to the $B$ deformation on the other hand are charged among the following symmetries have exactly the same charge pattern under as the states before but under the off diagonal U(1) combination:  $(U(1)^{1,3}$,$U(1)^{2,2},U(1)^{3,3})$.\\\\
Note that the R-symmetry of this model in eq.: \eqref{eq:RotoRcharge} is not corrected by non-Abelian gauge enhancements and hence the charges of all fermions is essentially given by the first twisted sector number. Hence the above described fermions have R-charge $q_R = 3$ and thus left-chiral super multiplets have R-charge $q_R (S) = 6$. Thus a VEV in any of those representation will break the R-symmetry in the 4D theory.
 A D-flat direction is given by the VEV configuration
\begin{align}
\label{eq:Aasinglets}
 A_a:& \qquad \langle S_{(a,a,a;1,2,3)} \rangle  = \langle S_{(a,a,a;2,3,1)}\rangle =\langle S_{(a,a,a;1,3,2)}\rangle = a_a \, , \\ 
 B:& \qquad \langle S_{(1,2,3;3,2,2)}\rangle  = \langle S_{(2,3,1;3,2,2)}\rangle =\langle S_{(2,1,3;3,2,2)} \rangle =  b \, ,
\end{align}
which is enforced by the triplet charge patter under three U(1)'s.
Indeed this VEV configuration is also F-flat as the above singlet fields always appear at most linear in any coupling. We observe that the VEV lies diagonally in the three Higgs fields whereas two of them become the Goldstone modes of the two broken $U(1)$'s and the third can be interpreted as the diagonal K\"{a}hler modulus in the orbifold. In addition we have explicitly checked that the three fields in eq. \eqref{eq:Aasinglets} take part in gauge invariant tree-linear couplings with exactly 27 pairs of fermions. Hence there are 54 fields that become massive by giving a non-trivial VEVs $a_a \neq 0$. These massive states originate from sectors $(k_0;k_1,k_2,k_3)$ and their conjugate ones
$(6-k_0 ;3- k_1 ,3- k_2 ,3- k_3)$ as the Higgs fields are untwisted sector fields.\\
To summarize we find that any Higgs VEV $a_a$ corresponding to the deformations $A_a$, breaks two $U(1)$'s and removes $54+2$ fermions which exactly matches our computation from the LGO side in Table~\ref{eq:SU(3)4LGOspectrum}. Switching on all deformations $A_a$ brings us to the orbifold point
where we find $84$ singlet states and two residual U(1)'s as well as three (not-necessarily independent) discrete symmetries. However we also find that the R-symmetry is broken as soon as we move towards finite volume geometries in the mirror.
\section{Summary and Discussion}
\label{sec:summary}
In this work we have classified the subset of heterotic (0,2) Landau-Ginzburg orbifolds with nine (2,2) superfields including their discrete quotients. In addition most of 
the computations in the literature do not go beyond the computation of the chiral ring or equivalently E$_6$ charged matter. We add the full computation of singlets and vector multiplets at the Fermat point that are uncharged under the generic E$_6$ to the existing literature. In total we find a set of 152 inequivalent models closed under mirror symmetry. Within this set we find models with $\mathcal{N}=4,2$ and $1$ supersymmetry as well as $\chi=0$
and $h^{1,1}=0$ models that do not have a geometric phase. Models with a geometric phase admit a $\mathbb{Z}_3^n$ orbifold phase where additional vector states at the Fermat point can be matched to the gauged isometries of the underlying orbifold lattice at the self-dual radius similar as in \cite{Beye:2014nxa}. Moreover we find hints that all models might live in a common moduli space as all of them satisfy a common relation \eqref{eq:relation} that relates all non-$E_6$ charged states with the amount of target space SUSY. In addition we extend the methods of \cite{Distler:1994hs} to compute discrete R-and non-R-symmetry 
charges of 4D states as well as their representations under the additional gauge symmetries in Section~\ref{subsec:TargetSymmetries}. By considering two examples we indeed find that there do not exist uncharged moduli fields in the spectrum.
\\\\
These methods pave the way for the second part of this work where we investigate non-Fermat deformations of Landau-Ginzburg models by considering two explicit examples. These deformations correspond to the world sheet K\"{a}hler parameters in the mirror dual geometry. Indeed we have constructed the two examples to be Landau-Ginzburg description of $\mathbb{Z}_3$ and $\mathbb{Z}_3 \times \mathbb{Z}_{3,\text{free}}$ orbifolds on non-factorized torus lattices.\\
By having full control over the spectrum and its symmetries we match the Landau-Ginzburg deformations perfectly to the effects of the Higgs effect in four dimensions. This formulation enables us to track the change in the spectrum and the breakdown of all symmetries through the various geometric phases. In particular for non factorized orbifolds these calculations have not been performed using the CFT techniques \cite{Bizet:2013wha} due to the loss of holomorphicity of the coordinates. Hence we provide a tool to compute discrete (non-)R-symmetries for non-factorized geometries. In the second example we can confirm the conjectured absence of a discrete R-symmetry in the $\mathbb{Z}_3 \times \mathbb{Z}_{3,\text{free}}$ as soon as one enters its large volume regime. In the first example however the R-symmetry is conserved as expected for an $\mathbb{Z}_3$ orbifold. Moreover this model admits very strong symmetries i.e. an additional $SU(3)^4 \rtimes S_4$ symmetry at the Fermat locus and an $E_6$ torus lattice in the mirror dual orbifold phase. 
 Due to these strong symmetries any further deformation corresponds to a simultaneous resolution of three singularities at once and all of them preserve the R-symmetry. Hence we provide the first example of a smooth Calabi-Yau geometry that admits a discrete R-symmetry. This conservation results from the very special locus in the K\"{a}hler moduli space where three fixed points have a common resolution divisor. In addition due to the $S_4$ permutation symmetry the K\"{a}hler parameters of the underlying torus and the resolution divisors are indistinguishable as also been observed in \cite{Blaszczyk:2011hs}.\\\\
The models we have considered in this work correspond to the heterotic standard embedding only and in the orbifold phase we always have variants of $\mathbb{Z}_3^n$ actions. Hence the next logical step would be to extend the above procedure to other examples i.e. prime and non-prime factor models and check if the models satisfy similar relations.\\
Finally it would be desirable to extend the above procedure also to (0,2) to models that can describe non-standard embedding models. However those attempts come genuinely with a lot more problems such as complications in the RG flow from the UV Landau-Ginzburg theory \cite{Bertolini:2014ela} and a much less well understood mirror map \cite{Melnikov:2012hk}.
\subsection*{Acknowledgments}
We would like to thank Hans Jockers and Thorsten Schimannek for useful discussions as well as Ralph Blumenhagen for references to earlier works.
This work is partially supported by the {\it Cluster of Excellence ‘Precision
Physics, Fundamental Interactions and Structure of Matter’} (PRISMA) DGF no.
EXC 1098, the DFG research grant HO 4166/2-1. 
The work of P.O. is partially supported by a scholarship of the Bonn-Cologne Graduate School BCGS, the SFB-Transregio TR33 The Dark Universe (Deutsche Forschungsgemeinschaft) and the European Union 7th network program Unification in the LHC era (PITN-GA-2009-237920). P.~O. would like to thank the KIAS for hospitality and financial support during the completion of this work.

\newpage
\appendix

\section{Additional Gaugino States  in $SU(3)^4$ Model}
We summarize the explicit form of the 32 additional gauginos of the $SU(3)^4$ model. Note that we find the 9 Cartan generators in the first twisted 
sector whereas the 24 roots are distributed throughout the other sectors
\begin{table}[h!]
\begin{center}
\renewcommand{\arraystretch}{1.3}
\begin{tabular}{|l|l|} \hline
Cartan Generators & $\left(\phi^{a,i}_{-\frac16} \overline\phi^{a,i}_{-\frac56} - 2 \psi^{a,i}_{-\frac23} \overline\psi^{a,i}_{-\frac13} \right)|1;0,0\rangle \,, a,i = 1,2,3$ \\ \hline
\multirow{6}*{$SU(3)^A$}& $ \overline{\phi}^{2,1}_{-\frac16} \overline{\phi}_{-\frac16}^{2,2} \overline{\phi}_{-\frac16}^{2,3} \overline{\psi}_0^{1,1} \overline{\psi}_0^{1,2} \overline{\psi}_0^{1,3} |1;1,0\rangle$ \\
&$ \overline{\phi}^{1,1}_{-\frac16} \overline{\phi}_{-\frac16}^{1,2} \overline{\phi}_{-\frac16}^{1,3} \overline{\psi}_0^{3,1} \overline{\psi}_0^{3,2} \overline{\psi}_0^{3,3} |3;1,0\rangle$ \\
&$ \overline{\phi}^{3,1}_{-\frac16} \overline{\phi}_{-\frac16}^{3,2} \overline{\phi}_{\frac16}^{3,3} \overline{\psi}_0^{2,1} \overline{\psi}_0^{2,2} \overline{\psi}_0^{2,3} |5;1,0\rangle$ \\
&$ \overline{\phi}^{1,1}_{-\frac16} \overline{\phi}_{-\frac16}^{1,2} \overline{\phi}_{-\frac16}^{1,3} \overline{\psi}_0^{1,1} \overline{\psi}_0^{1,2} \overline{\psi}_0^{1,3} |1;2,0\rangle$ \\
&$ \overline{\phi}^{2,1}_{-\frac16} \overline{\phi}_{-\frac16}^{2,2} \overline{\phi}_{-\frac16}^{2,3} \overline{\psi}_0^{3,1} \overline{\psi}_0^{3,2} \overline{\psi}_0^{3,3} |3;2,0\rangle$ \\
&$ \overline{\phi}^{3,1}_{-\frac16} \overline{\phi}_{-\frac16}^{3,2} \overline{\phi}_{-\frac16}^{3,3} \overline{\psi}_0^{1,1} \overline{\psi}_0^{1,2} \overline{\psi}_0^{1,3} |5;2,0\rangle$ \\ \hline

\multirow{6 }*{$SU(3)^B$}& $\overline{\phi}^{1,3}_{-\frac16} \overline{\phi}_{-\frac16}^{2,3} \overline{\phi}_{-\frac16}^{3,3} \overline{\psi}_0^{1,2} \overline{\psi}_0^{2,2} \overline{\psi}_0^{3,2} |1;0,1\rangle$ \\
&$ \overline{\phi}^{1,2}_{-\frac16} \overline{\phi}_{-\frac16}^{2,2} \overline{\phi}_{-\frac16}^{3,2} \overline{\psi}_0^{1,1} \overline{\psi}_0^{2,1} \overline{\psi}_0^{3,1} |3;0,1\rangle$ \\
&$ \overline{\phi}^{1,1}_{-\frac16} \overline{\phi}_{-\frac16}^{2,1} \overline{\phi}_{-\frac16}^{3,1} \overline{\psi}_0^{1,3} \overline{\psi}_0^{2,3} \overline{\psi}_0^{3,3} |5;0,1\rangle$ \\
&  $  \overline{\phi}^{1,2}_{-\frac16} \overline{\phi}_{-\frac16}^{2,2} \overline{\phi}_{-\frac16}^{3,2} \overline{\psi}_0^{1,3} \overline{\psi}_0^{2,3} \overline{\psi}_0^{3,3}  |1;0,2\rangle$ \\
&$ \overline{\phi}^{1,3}_{-\frac16} \overline{\phi}_{-\frac16}^{2,3} \overline{\phi}_{-\frac16}^{3,3} \overline{\psi}_0^{1,1} \overline{\psi}_0^{2,1} \overline{\psi}_0^{3,1}|3;0,2\rangle$ \\
&$ \overline{\phi}^{1,1}_{-\frac16} \overline{\phi}_{-\frac16}^{2,1} \overline{\phi}_{-\frac16}^{3,1} \overline{\psi}_0^{1,2} \overline{\psi}_0^{2,2} \overline{\psi}_0^{3,2}|5;0,2\rangle$ \\ \hline

\multirow{6 }*{$SU(3)^C$}&$ \overline{\phi}^{1,2}_{-\frac16} \overline{\phi}_{-\frac16}^{1,1} \overline{\phi}_{-\frac16}^{3,3} \overline{\psi}_0^{1,1} \overline{\psi}_0^{2,3} \overline{\psi}_0^{3,2} |1;1,1\rangle$ \\
&$ \overline{\phi}^{1,1}_{-\frac16} \overline{\phi}_{-\frac16}^{2,3} \overline{\phi}_{-\frac16}^{3,2} \overline{\psi}_0^{1,3} \overline{\psi}_0^{2,2} \overline{\psi}_0^{3,2} |3;1,1\rangle$ \\
&$ \overline{\phi}^{1,3}_{-\frac16} \overline{\phi}_{-\frac16}^{2,2} \overline{\phi}_{-\frac16}^{3,1} \overline{\psi}_0^{1,2} \overline{\psi}_0^{2,1} \overline{\psi}_0^{3,3} |5;1,1\rangle$ \\
&$ \overline{\phi}^{1,1}_{-\frac16} \overline{\phi}_{-\frac16}^{2,3} \overline{\phi}_{-\frac16}^{3,2} \overline{\psi}_0^{1,2} \overline{\psi}_0^{2,1} \overline{\psi}_0^{3,3} |1;2,2\rangle$ \\
&$ \overline{\phi}^{1,2}_{-\frac16} \overline{\phi}_{-\frac16}^{2,1} \overline{\phi}_{-\frac16}^{3,3} \overline{\psi}_0^{1,3} \overline{\psi}_0^{2,2} \overline{\psi}_0^{3,1} |3;2,2\rangle$ \\
&$ \overline{\phi}^{1,3}_{-\frac16} \overline{\phi}_{-\frac16}^{2,2} \overline{\phi}_{-\frac16}^{3,1} \overline{\psi}_0^{1,1} \overline{\psi}_0^{2,3} \overline{\psi}_0^{3,2} |5;2,2\rangle$ \\ \hline

\multirow{6 }*{$SU(3)^D$}&$ \overline{\phi}^{1,3}_{-\frac16} \overline{\phi}_{-\frac16}^{1,1} \overline{\phi}_{-\frac16}^{3,2} \overline{\psi}_0^{1,1} \overline{\psi}_0^{2,2} \overline{\psi}_0^{3,3} |1;1,2\rangle$ \\
&$ \overline{\phi}^{1,1}_{-\frac16} \overline{\phi}_{-\frac16}^{2,2} \overline{\phi}_{-\frac16}^{3,3} \overline{\psi}_0^{1,2} \overline{\psi}_0^{2,3} \overline{\psi}_0^{3,1} |3;1,2\rangle$ \\
&$ \overline{\phi}^{1,2}_{-\frac16} \overline{\phi}_{-\frac16}^{2,3} \overline{\phi}_{-\frac16}^{3,1} \overline{\psi}_0^{1,3} \overline{\psi}_0^{2,1} \overline{\psi}_0^{3,2} |5;1,2\rangle$ \\
&$ \overline{\phi}^{1,1}_{-\frac16} \overline{\phi}_{-\frac16}^{2,2} \overline{\phi}_{-\frac16}^{3,3} \overline{\psi}_0^{1,3} \overline{\psi}_0^{2,1} \overline{\psi}_0^{3,2} |1;2,1\rangle$ \\
&$ \overline{\phi}^{1,3}_{-\frac16} \overline{\phi}_{-\frac16}^{2,1} \overline{\phi}_{-\frac16}^{3,2} \overline{\psi}_0^{1,2} \overline{\psi}_0^{2,3} \overline{\psi}_0^{3,1} |3;2,1\rangle$ \\
&$ \overline{\phi}^{1,2}_{-\frac16} \overline{\phi}_{-\frac16}^{2,3} \overline{\phi}_{-\frac16}^{3,1} \overline{\psi}_0^{1,1} \overline{\psi}_0^{2,2} \overline{\psi}_0^{3,3} |5;2,1\rangle$ \\ \hline
\end{tabular}
\caption{\label{tab:SU34Vectors}All 32 gaugino states that belong to the $SU(3)^4$ model. We have collected the roots in groups of their SU(3) roots.   }
\end{center}
\end{table}

\section{Mirror dual GLSM Descriptions}
In this section we give the explicit charge assignments for the GLSMs that give the geometric description of the two mirror-dual LGO models that we consider.
\subsection{GLSM with $\mathbb{Z}_3$ Orbifold Phase on E$_6$ Lattice}
\label{app:SU(3)4GLSM}
The GLSM specified in Table \ref{tab:SU(3)4GLSM} admits a $\mathbb{Z}_3$ orbifold phase with an E$_6$ torus lattice. The orbifold blow-up cycles are controlled by the exceptional coordinates $E_{1,2,3}$ and their gaugings. The orbifold phase is obtained by sending their FI terms to negative values. Hence each cycle controls the size of 9 blow-up modes. Setting the FI parameters of torus divisors $U(1)_{1,2,3}$ to negative values results in the LGO phase.
\begin{table}[h!]
\begin{center}
\begin{tabular}{c| c c c| c c c| c c c|c c c | c c c|}
 & $\Phi^{1,1}$ & $\Phi^{1,2}$ & $\Phi^{1,3}$ & $\Phi^{2,1}$ & $\Phi^{2,2}$ & $\Phi^{2,3}$ & $\Phi^{3,1}$  & $\Phi^{3,2}$ & $\Phi^{3,3}$ & $C_1$ & $C_2$ & $C_3$ & $E_1$ & $E_2$ & $E_3$ \\ \hline  
$\rm{U}(1)_R$ & 0 & 0 & 0 & 0 & 0 & 0 & 0 & 0 & 0 & 1 & 1 & 1 & 0 & 0 & 0  \\ \hline
$\rm{U}(1)_1$ & 1 & 1 & 1 & 0 & 0 & 0 & 0 & 0 & 0 & -3 & 0 & 0 & 0 & 0 & 0 \\ 
$\rm{U}(1)_2$ & 0 & 0 & 0 & 1 & 1 & 1 & 0 &0 & 0& 0 & -3 & 0 & 0 & 0 & 0 \\
$\rm{U}(1)_3$ & 0 & 0 & 0 & 0 & 0 & 0 & 1 & 1 & 1 & 0 & 0 & -3 & 0 & 0 & 0 \\ \hline
$\rm{U}(1)_{E_1}$ & 1 & 0 & 0 & 1 & 0 & 0 & 1 & 0 & 0 & 0 & 0 & 0 & -3  & 0 & 0\\
$\rm{U}(1)_{E_2}$ & 0 & 1 & 0 & 0 & 1 & 0 & 0 & 1 & 0 & 0 & 0 & 0 & 0  & -3 & 0\\
$\rm{U}(1)_{E_3}$ & 0 & 0 & 1 & 0 & 0 & 1 & 0 & 0 & 1 & 0 & 0 & 0 & 0  & 0 & -3\\ \hline
\end{tabular}
\caption{\label{tab:SU(3)4GLSM}The charge assignment of a GLSM with $\mathbb{Z}_3 $ orbifold phase with E$_6$ torus structure.}
\end{center}
\end{table}

\subsection{GLSM with $\mathbb{Z}_3 \times \mathbb{Z}_{3,\text{free}}$ Orbifold Phase}
\label{app:rotoglsm}
In Table \ref{tab:RotoGLSM} we give the GLSM of the $\mathbb{Z}_3$ orbifold that has an additional freely acting $\mathbb{Z}_{3,\text{free}}$ modded out, realized 
by the last column. The orbifold phase is obtained by setting the FI term of the exceptional $\rm{U}(1)_E$ to negative values. The LGO phase we obtain by
setting all other FI terms to negative values as well. We note that we can not unhiggs the $\mathbb{Z}_3$ symmetry at the GLSM level to a U(1) symmetry on the world sheet.
\begin{table}[h!]
\begin{center}
\begin{tabular}{c| c c c| c c c| c c c|c c c | c|}
 & $\Phi_{1}$ & $\Phi_{2}$ & $\Phi_{3}$ & $\Phi_{4}$ & $\Phi_{5}$ & $\Phi_{6}$ & $\Phi_{7}$  & $\Phi_{8}$ & $\Phi_{9}$ & $C_1$ & $C_2$ & $C_3$ & $E_1$ \\ \hline  
$\rm{U}(1)_R$ & 0 & 0 & 0 & 0 & 0 & 0 & 0 & 0 & 0 & 1 & 1 & 1 & 0 \\ \hline
$\rm{U}(1)_1$ & 1 & 1 & 1 & 0 & 0 & 0 & 0 & 0 & 0 & -3 & 0 & 0 & 0 \\ 
$\rm{U}(1)_2$ & 0 & 0 & 0 & 1 & 1 & 1 & 0 &0 & 0& 0 & -3 & 0 & 0 \\
$\rm{U}(1)_3$ & 0 & 0 & 0 & 0 & 0 & 0 & 1 & 1 & 1 & 0 & 0 & -3 & 0 \\ \hline
$\rm{U}(1)_E$ & 1 & 0 & 0 & 1 & 0 & 0 & 1 & 0 & 0 & 0 & 0 & 0 & -3 \\ \hline
$\mathbb{Z}_{3,\text{free}}$ & 0 & 1 & 0 & 0 & 2 & 0 & 1 & 1 & 1 & 0 & 0 & 0 & 0 \\ \hline
\end{tabular}
\caption{\label{tab:RotoGLSM}The charge assignment of a GLSM with $\mathbb{Z}_3 \times \mathbb{Z}_{3,\text{free}}$ orbifold phase.}
\end{center}
\end{table}
\section{List of Charges of $A_1^9$ Classification}
\label{app:LGOList}
\begin{figure}[h!]
\begin{picture}(0,0)
\put(0,-350){
{\tiny 
\renewcommand{\arraystretch}{0.8}
\begin{tabular}{|cccc|c|c|c|c|c|c|} \cline{1-4}
 S & $27$ & $\overline{27}$ & Adj.  &  \multicolumn{6}{|c}{} \\ \hline
\multicolumn{10}{|c|}{\vphantom{\large|}\bf $\mathcal{N}=4$} \\ 
\hline
 32 & 3 & 3 & 1 &(1,1,1,0,0,0,0,0,0) & (1,1,1,0,0,0,0,0,0) & (1,1,1,0,0,0,0,0,0) &\multicolumn{3}{|c}{}\\
 96 & 9 & 9 & 3 & (0,0,0,1,1,1,0,0,0) & (0,0,0,1,1,1,0,0,0) & (0,0,0,1,1,1,0,0,0)&\multicolumn{3}{|c}{}\\
 &  &  &  & & (0,0,0,0,0,0,1,-1,0) & (0,0,0,0,1,-1,1,-1,0)&\multicolumn{3}{|c}{} \\
\cline{1-7}
86 & 3 & 3 & 1 &  (1,1,1,0,0,0,0,0,0)  & \multicolumn{5}{|c}{}\\
258 & 9 & 9 & 3 & (0,0,0,1,1,1,0,0,0) & \multicolumn{5}{|c}{}\\
  &  &  &  & (1,-1,0,1,-1,0,1,1,-1,0)  & \multicolumn{5}{|c}{}\\
\hline 
\multicolumn{10}{|c|}{{\vphantom{\large|}\bf $\mathcal{N}=2$}} \\ \hline
 14 & 1 & 1 & 1 & (1,1,1,0,0,0,0,0,0)&  (1,2,0,0,0,0,0,0,0) & (1,2,0,1,2,0,0,0,0) & (1,2,0,0,0,0,0,0,0) & \multicolumn{2}{|c}{} \\
 194 & 21 & 21 & 1 &  & (0,0,0,1,1,1,0,0,0) & (0,0,0,0,0,0,1,1,1) &  (0,0,1,1,1,1,1,1,1)    & \multicolumn{2}{|c}{} \\
\cline{1-8}
 14 & 1 & 1 & 1 &  (1,1,1,0,0,0,0,0,0)& (1,2,0,0,0,0,0,0,0) & (1,2,0,0,0,0,0,0,0) & (1,2,0,1,2,0,0,0,0) &\multicolumn{2}{|c}{}\\
 194 & 21 & 21 & 1 &  (0,0,1,1,1,0,0,0,0)    &  (0,0,0,1,2,0,0,0,0) &  (0,2,1,0,0,0,0,0,0) &  (0,2,1,1,2,0,0,0,0)  &\multicolumn{2}{|c}{} \\
  &  &  &  &   (0,0,0,0,0,1,1,1,0) &    (0,0,0,0,0,0,1,1,1) &       (0,0,0,1,2,0,0,0,0)  & (0,0,0,0,0,0,1,2,0)      &\multicolumn{2}{|c}{}    \\
\cline{1-8}
 14 & 1 & 1 & 1 & (1,2,0,1,1,1,0,0,0) & (1,1,1,0,0,0,0,0,0)& (1,1,1,1,2,0,0,0,0) & (1,2,0,1,2,0,0,0,0) &\multicolumn{2}{|c}{} \\
194 & 3 & 3 & 1 & (0,0,0,1,1,1,1,1,1) & (0,0,0,1,1,1,0,0,0)& (0,0,0,1,1,1,0,0,0)  & (0,2,1,0,2,1,0,0,0)&\multicolumn{2}{|c}{}  \\
  &  &  &       &    & (0,0,1,1,1,0,1,2,0)   &  (0,0,0,0,0,0,1,2,0) &  (1,1,1,0,0,0,1,2,0) &\multicolumn{2}{|c}{}  \\
\cline{1-8}

 14 & 1 & 1 & 1 & (1,1,1,0,0,0,0,0,0) & (1,2,0,1,2,0,0,0,0)& (1,2,0,1,2,0,0,0,0) & \multicolumn{3}{|c}{} \\
194 & 9 & 9 & 1 & (1,2,0,1,2,0,1,2,0) & (0,1,2,0,0,0,0,0,0)& (0,2,1,0,2,1,0,0,0)  & \multicolumn{3}{|c}{} \\
 &  &  &       &  &   (0,1,0,1,1,0,1,2,0)   &  (1,1,0,0,1,0,1,2,0) &   \multicolumn{3}{|c}{} \\
\cline{1-7}

 32 & 1 & 1 & 1 & (1,1,1,0,0,0,0,0,0) & (1,2,0,1,2,0,0,0,0)& (1,2,0,1,2,0,0,0,0) & \multicolumn{3}{|c}{}\\
248 & 9 & 9 & 1 & (1,2,0,1,2,0,0,0,0) & (0,0,0,1,1,1,0,0,0)& (1,1,1,0,0,0,0,0,0)  & \multicolumn{3}{|c}{}\\
 &  &  &       & &    (0,0,0,0,0,0,1,2,0)   &  (0,0,0,2,1,0,1,2,0) &  \multicolumn{3}{|c}{} \\
\hline
\multicolumn{10}{|c|}{{\vphantom{\large|}\bf $\mathcal{N}=1$, $\chi=0$}} \\ \hline

 8 & 0 & 0 & 1 & (1,1,1,1,2,0,0,0,0) &  (1,2,0,0,0,0,0,0,0) & (1,2,0,0,0,0,0,0,0) & (1,2,0,0,0,0,0,0,0)  &\multicolumn{2}{|c}{} \\
 252& 13 & 13 & 0  & (0,1,0,0,0,0,1,1,0) & (0,1,0,0,1,0,0,1,0) & (0,1,0,1,1,0,1,2,0) & (0,2,1,0,0,0,0,0,0) &\multicolumn{2}{|c}{}  \\
 &  &  &  & &   (0,0,0,1,1,1,0,0,0) &    (0,0,0,1,1,1,0,0,0) &       (0,1,0,1,1,0,1,2,0)  &\multicolumn{2}{|c}{}       \\
\cline{1-8} 
 8 & 0 & 0 & 1 & (1,2,0,1,2,0,0,0,0) &  \multicolumn{5}{|c}{}    \\
 252& 9 & 9 & 0  & (0,1,0,0,1,1,1,2,0) &  \multicolumn{5}{|c}{}  \\
 &  &  &  & (0,0,0,0,1,0,0,1,1)&  \multicolumn{5}{|c}{}        \\
\cline{1-5}
 14 & 0 & 0 & 1 & (1,1,1,1,2,0,0,0,0) &  \multicolumn{5}{|c}{} \\
 270& 9 & 9 & 0  & (0,1,0,2,1,0,1,1,0) & \multicolumn{5}{|c}{}  \\
 &  &  &  & (1,0,0,0,0,0,0,1,1)&    \multicolumn{5}{|c}{}       \\
\cline{1-5}
8 & 0 & 0 & 1 & (1,2,0,1,2,0,0,0,0) & \multicolumn{5}{|c}{}  \\
 252& 7 & 7 & 0  & (2,1,0,0,0,0,1,1,1) & \multicolumn{5}{|c}{}  \\
  &  &  &  & (1,0,0,0,0,1,0,0,1)&   \multicolumn{5}{|c}{}        \\
\hline 

\multicolumn{10}{|c|}{{\vphantom{\large|}\bf $\mathcal{N}=1$ Mirror Pairs}} \\ \hline
 8  & 0  & 0 & 1 & (0,0,0,0,0,0,0,0,0) &  (1,2,0,0,0,0,0,0,0) & (1,2,0,0,0,0,0,0,0) & (1,2,0,0,0,0,0,0,0) &\multicolumn{2}{|c}{}  \\
 252& 84 & 0 & 0 &  &  & (0,0,0,1,2,0,0,0,0) & (0,0,0,1,2,0,0,0,0) &\multicolumn{2}{|c}{}  \\
    &    &   &   &  &   &    & (0,0,0,0,0,0,1,2,0) &\multicolumn{2}{|c}{}        \\ 
\cline{1-8}
 8 & 0 & 0 & 1 & (1,1,1,0,0,0,0,0,0) &   \multicolumn{5}{|c}{}  \\
 252& 0 & 84 & 0  & (0,1,0,0,1,1,0,0,0) &  \multicolumn{5}{|c}{}    \\
 &  &  &  & (0,1,0,0,0,0,0,1,1)&    \multicolumn{5}{|c}{}       \\
\hline

 8 & 0 & 0 & 1 & (1,1,1,1,2,0,0,0,0) & (1,1,1,1,1,1,0,0,0)  & (1,2,0,1,2,0,0,0,0)   &   (1,2,0,0,0,0,0,0,0) & (1,0,0,1,1,1,1,1,0)  & (1,2,0,1,2,0,0,0,0)          \\
 252& 40 & 4 & 0  &  & (0,0,0,0,0,1,1,1,0)  &  (0,0,0,0,0,0,1,2,0) & (0,2,1,1,1,1,0,0,0) &  (0,1,0,1,1,1,1,1,0)&  (0,0,0,0,0,0,1,2,0)         \\
 &  &  &  & &    &     &  (0,0,0,0,0,1,1,1,0) &  (0,0,1,1,1,1,1,1,0)&    (0,0,0,0,0,2,0,0,1)          \\
\hline 

8 & 0 & 0 & 1 & (1,1,1,0,0,0,0,0,0) & (1,2,0,0,0,0,0,0,0)   &  (0,0,1,1,1,0,0,0,0) &   \multicolumn{3}{|c}{} \\
 252& 4 & 40 & 0  & (0,1,0,1,1,0,0,0,0) &  (0,0,0,1,1,1,0,0,0)      & (0,1,0,1,1,0,0,0,0) &   \multicolumn{3}{|c}{} \\
  &  &  &  & &   (0,0,0,0,1,0,1,1,0)               &   (1,0,0,1,1,0,0,0,0)  &        \multicolumn{3}{|c}{}        \\
\cline{1-9}

 14 & 0 & 0 & 1 & (1,2,0,1,2,0,0,0,0) & (1,2,0,1,2,0,1,2,0)  & (1,2,0,1,2,0,0,0,0)   &   (1,2,0,0,0,0,0,0,0) & (1,2,0,1,2,0,0,0,0) &   \multicolumn{1}{|c}{}   \\
 270& 36 & 0 & 0  &  & (0,0,1,0,0,2,0,0,0)  &  (2,1,0,0,1,2,1,2,0) &                     (0,1,0,1,1,0,1,2,0) &  (2,1,0,1,2,0,0,0,0)&    \multicolumn{1}{|c}{}    \\
 &  &  &  & &    &     &                                                                 (0,0,1,0,0,2,0,0,0) &  (0,0,0,0,0,0,1,2,0)&       \multicolumn{1}{|c}{}       \\
\cline{1-9}

14 & 0 & 0 & 1 & (1,1,1,0,0,0,0,0,0) &  \multicolumn{5}{|c}{}  \\
 270& 0 & 36 & 0  & (0,1,0,1,1,0,0,0,0) &    \multicolumn{5}{|c}{}\\
  &  &  &  & (0,0,0,0,1,0,1,1,0)  &          \multicolumn{5}{|c}{}     \\
\cline{1-6}

 32 & 0 & 0 & 1 & (1,2,0,1,2,0,1,2,0) & (1,2,0,1,2,0,0,0,0)  &   \multicolumn{4}{|c}{}     \\
324 & 36 & 0 & 0  & (2,1,2,2,1,0,0,0,1) & (0,2,0,0,0,1,1,2,0) &  \multicolumn{4}{|c}{}    \\
 &  &  &  & &  (0,1,2,0,0,2,0,0,1)  &  \multicolumn{4}{|c}{}           \\
\cline{1-6}

32 & 0 & 0 & 1 & (1,0,0,0,2,1,1,1,0)&   \multicolumn{5}{|c}{}  \\
 324& 0 & 36 & 0 & (0,1,0,1,1,1,0,2,0) &     \multicolumn{5}{|c}{}  \\
  &  &  &  &   (0,0,1,2,1,1,1,0,0)     &      \multicolumn{5}{|c}{}       \\
\cline{1-9}

 8 & 0 & 0 & 1 & (1,1,1,2,2,2,0,0,0) &  (1,2,0,1,1,1,0,0,0)   & (1,2,0,0,0,0,0,0,0)  &  (1,2,0,1,2,0,0,0,0)  & (1,2,0,0,0,0,0,0,0)          &  \multicolumn{1}{|c}{}  \\
252 & 24 & 12 & 0  &  &  (0,0,0,0,0,0,1,2,0)  &  (1,0,2,1,2,0,0,0,0) &     (0,1,2,0,0,0,0,0,0)                  &  (0,1,1,1,0,0,0,0,0) &    \multicolumn{1}{|c}{}     \\
 &  &  &  & &     &  (0,0,0,0,0,0,1,2,0)  & (0,1,1,0,1,0,1,2,0)    & (0,0,0,2,1,0,0,0,0) &         \multicolumn{1}{|c}{}     \\
\cline{1-9}

8 & 0 & 0 & 1 & (1,2,0,1,2,0,0,0,0)&  (1,2,0,1,2,0,0,0,0) &   \multicolumn{4}{|c}{}   \\
 252& 12 & 24 & 0  & (0,1,0,0,1,1,2,1,0) &  (0,2,1,0,2,1,0,0,0)     & \multicolumn{4}{|c}{}   \\
  &  &  &  & &   (0,0,0,0,0,1,0,1,1)    & \multicolumn{4}{|c}{}     \\
\cline{1-6}

8 & 0 & 0 & 1 & (1,2,0,0,0,0,0,0,0)&  (1,2,0,1,2,0,0,0,0) &  \multicolumn{4}{|c}{}   \\
 252& 18 & 6 & 0  & (0,0,0,1,1,1,1,2,0) &  (0,1,0,0,1,0,0,1,2)     & \multicolumn{4}{|c}{}   \\
  &  &  &  & &   (0,0,1,0,0,2,0,0,0)     &  \multicolumn{4}{|c}{}         \\
\cline{1-6}

 8 & 0 & 0 & 1 & (1,2,0,1,2,0,0,0,0)  &    \multicolumn{5}{|c}{}    \\
252 & 6 & 18 & 0  & (0,0,1,1,0,1,0,0,0)  &  \multicolumn{5}{|c}{}      \\
 &  &  &  & (0,0,0,0,0,1,1,1,0)  &   \multicolumn{5}{|c}{}           \\
\cline{1-7}

14 & 0 & 0 & 1 & (1,0,0,0,1,2,2,0,0)&  (1,0,0,0,0,2,2,1,0) &  (1,0,0,0,2,1,1,1,0)  &   \multicolumn{3}{|c}{}    \\
 270& 18 & 6 & 0  & (0,1,2,2,1,0,0,0,0) &  (0,1,0,0,2,1,1,1,0)     & (0,1,0,0,1,1,1,2,0)&   \multicolumn{3}{|c}{}    \\
  &  &  &  & &   (0,0,1,2,0,0,0,0,0)     & (0,0,1,2,1,1,1,0,0)  &          \multicolumn{3}{|c}{}         \\
\cline{1-7}

 14 & 0 & 0 & 1 & (1,0,0,1,1,0,2,1,0)  & (1,0,0,0,1,1,2,1,0)   &  (1,0,0,0,2,2,0,1,0)  &  \multicolumn{3}{|c}{}    \\
270 & 6 & 18 & 0  & (0,1,2,1,1,1,0,0,0)  & (0,1,0,1,0,1,0,0,0)  &  (0,1,0,0,0,1,0,2,0)  &                   \multicolumn{3}{|c}{}     \\
 &  &  &  &   &   (0,0,1,1,1,0,0,0,0)  &  (0,0,1,2,1,1,1,0,0)  & \multicolumn{3}{|c}{}         \\
\cline{1-7}

8 & 0 & 0 & 1 & (1,0,0,2,1,0,2,0,0)&  (1,0,0,0,2,1,1,1,0) &   \multicolumn{4}{|c}{}  \\
 252 & 16 & 4 & 0  & (0,1,2,1,1,1,0,0,0) &  (0,1,0,0,0,0,0,2,0)     & \multicolumn{4}{|c}{}  \\
  &  &  &  & &   (0,0,1,2,1,1,1,0,0)     &   \multicolumn{4}{|c}{}      \\
\cline{1-6}

8 & 0 & 0 & 1 & (1,0,0,0,2,2,1,0,0)  & (1,0,0,0,0,2,1,2,0)   &    \multicolumn{4}{|c}{}  \\
252 & 4 & 16 & 0  & (0,1,0,1,0,1,0,0,0)  & (0,1,0,0,2,1,1,1,0)  &    \multicolumn{4}{|c}{}    \\
 &  &  &  &  (0,0,1,1,1,0,0,0,0) &   (0,0,1,2,1,1,1,0,0)  &   \multicolumn{4}{|c}{}           \\
\cline{1-6}

8 & 0 & 0 & 1 & (1,0,0,0,0,2,2,1,0)&   \multicolumn{5}{|c}{}     \\
 252 & 27 & 3 & 0 & (0,1,0,1,0,2,1,1,0) &    \multicolumn{5}{|c}{}  \\
  &  &  &  &(0,0,1,1,1,0,0,0,0) &         \multicolumn{5}{|c}{}      \\
\cline{1-5}

8 & 0 & 0 & 1 & (1,0,0,0,1,1,0,0,0)  &   \multicolumn{5}{|c}{}    \\
252 & 3 & 27 & 0  & (0,1,0,1,0,1,0,0,0)  &   \multicolumn{5}{|c}{}     \\
 &  &  &  &  (0,0,1,1,1,0,0,0,0) &      \multicolumn{5}{|c}{}          \\
\cline{1-5}

8 & 0 & 0 & 1 & (1,0,0,0,2,1,0,2,0)  &    \multicolumn{5}{|c}{}   \\
252 & 12 & 0 & 0  & (0,1,0,0,1,1,1,2,0)  &    \multicolumn{5}{|c}{}    \\
 &  &  &  &  (0,0,1,2,1,1,1,0,0) &    \multicolumn{5}{|c}{}          \\
\cline{1-5}

\end{tabular}
}
}
\end{picture}
\caption{\label{fig:LGOmodels}Charge assignment and matter content for all $A^9_1$ Fermat quotients.}
\end{figure}
\newpage
In this appendix we give the full classification of $A_1^9$ Gepner models and their discrete quotients.\\In Table~\ref{fig:LGOmodels} we list the charge vectors of the nine chiral superfields under the respective $\mathbb{Z}_3$ discrete symmetry. For convenience we only give the charge vectors for half of the models whereas the other half can be constructed from the mirror dual charge assignment \eqref{eq:mirrormap}. The first four rows highlight the 4D representations under E$_6$. In the first column we give chiral fermions with a left-chiral super multiplet. In the second one we give gauginos within a super vector multiplet.
 Note that adjoint valued chiral fermions signal higher SUSY models.
\clearpage
\providecommand{\href}[2]{#2}\begingroup\raggedright\endgroup

\end{document}